\documentclass[a4paper,fleqn]{cas-sc}

\usepackage[authoryear,longnamesfirst]{natbib}
\usepackage{amssymb}
\usepackage{amsmath}
\usepackage{url}
\usepackage{thmtools}    % 核心宏包，用于复述定理
\usepackage{amsthm}
\usepackage{enumitem}
\usepackage{algorithm}
\usepackage{algorithmic}
\usepackage[switch]{lineno}
\usepackage[utf8]{inputenc}
\usepackage{caption}
\usepackage{graphicx}
\usepackage{bm}
\usepackage{multirow}
\usepackage{threeparttable}

\newtheorem{property}{Property}
\newtheorem{definition}{Definition}
\def\tsc#1{\csdef{#1}{\textsc{\lowercase{#1}}\xspace}}
\tsc{WGM}
\tsc{QE}
\tsc{EP}
\tsc{PMS}
\tsc{BEC}
\tsc{DE}
\begin{document}
\let\WriteBookmarks\relax
\def\floatpagepagefraction{1}
\def\textpagefraction{.001}

\shorttitle{}

\shortauthors{ et~al.}

% ----------------- Authors -----------------

\author[1]{Yansong Wang}
\ead{yansong0682@email.swu.edu.cn}

\author[1]{Qisen Chai}
\ead{cqsllyt@email.swu.edu.cn}

\author[1]{Junjie Huang}
\ead{junjiehuang.cs@outlook.com}

\author[1,2]{Tao Jia}
\cormark[1]
\ead{tjia@swu.edu.cn}

% ---------------- Affiliations ----------------

\affiliation[1]{
    organization={College of Computer and Information Science,
                  Southwest University},
    city={Chongqing},
    postcode={400715},
    country={China}
}

\affiliation[2]{
    organization={College of Computer and Information Science,
                  Chongqing Normal University},
    city={Chongqing},
    postcode={401331},
    country={China}
}

% -------------- Corresponding Author ---------

\cortext[1]{Corresponding author at:
College of Computer and Information Science,
Southwest University, Chongqing, 400715, China.}

\title [mode = title]{CasHyper: Modeling Multi-Factor Interactions and Asymmetric Influence in Cascade Prediction via Hypercomplex Algebra}

\begin{abstract}
Cascade prediction aims to forecast the future evolution of information diffusion and is an important task in social network analysis. Most existing methods represent users and cascades in real-valued Euclidean spaces, which may offer limited capacity to model multi-factor interactions and asymmetric influence in complex diffusion processes. To address these challenges, we propose CasHyper, a hypercomplex-based cascade prediction framework that integrates hypercomplex algebra with graph neural networks and Transformers. CasHyper organizes user and cascade embeddings into multiple hypercomplex components,  with each component encoding distinct latent factors, and uses structured algebraic operations to model cross-component interactions. The non-commutative hypercomplex product further yields order-dependent interaction representations, enabling the model to learn asymmetric influence among users.
In the network encoding stage, a static heterogeneous graph channel captures user-user and user-cascade relations, while a Neural ODE-based dynamic channel models the continuous evolution of node states and edge weights. A hypercomplex Transformer with norm-preserving rotary transformations is then introduced to incorporate relative positional information into cascade sequence modeling. Finally, an interaction-aware module and a linear projection module are combined to estimate the probability of each candidate user participating next. Experiments on four real-world datasets show that CasHyper achieves competitive or better performance than the compared methods under the evaluated settings, including relative improvements of up to $43.24\%$ on Android and $18.43\%$ on Christianity in Hits@50. Additional analyses of architectural variants, graph incompleteness, computational efficiency, and different Cayley-Dickson algebras further demonstrate the empirical effectiveness of the proposed framework and reveal dataset-dependent behaviors of different hypercomplex structures. We release the code at \url{https://github.com/MrYansong/CasHyper}.
\end{abstract}

% \includegraphics{figs/grabs.pdf}

% \begin{highlights}
% \item Proposes CasHyper for hypercomplex-based information cascade prediction.
% \item Models multi-factor interactions and asymmetric influence via hypercomplex algebra.
% \item Introduces a hypercomplex Transformer with rotary positional encoding.
% \item Explores the impact of different hypercomplex algebras via Cayley-Dickson construction.
% \end{highlights}

% Keywords

\maketitle

\section{Introduction}
In the realm of interconnected social networks, information cascades represent the dynamic process by which information propagates from one user to another. Cascade prediction aims to forecast the future evolution of cascades by analyzing early-stage diffusion data. This research holds significant practical value for social recommendation \citep{wang2025generative, amran2026muse}, public opinion governance \citep{zhang2024predicting, yu2026opinion}, and viral marketing \citep{tu2022viral}, while also offering profound insights into the underlying information diffusion mechanisms.

Cascade prediction encompasses tasks such as popularity prediction \citep{wang2022casseqgcn,sang2025dyhgtcr}, outbreak prediction \citep{cui2013cascading}, and retweeting behavior prediction \citep{qiao2023rotdiff,liu2025gthnn}. Early studies mainly relied on diffusion models (e.g., Independent Cascade \citep{goldenberg2001talk} and Linear Threshold \citep{kempe2003maximizing}) or feature-engineered machine learning methods \citep{shulman2016predictability}. These methods, valued for their simplicity, were widely adopted but often suffered from limited generalization due to the numerous confounding factors in information propagation. The advent of network embedding techniques marked a critical breakthrough \citep{grover2016node2vec}, as they effectively encode topological structures into low-dimensional representations. Building upon this foundation, graph neural networks and temporal modeling methods \citep{wang2022ccasgnn, cheng2023enhancing} further enhance the modeling of nonlinear user interactions and dynamic diffusion processes, leading to substantial performance gains by jointly capturing structural, temporal, and content information \citep{qiao2023rotdiff,he2025pattern}.
A key limitation of most existing methods is their reliance on real-valued Euclidean embeddings with symmetric similarity functions (e.g., dot-product) to model user interactions. 
Recent studies have uncovered that information spreading dynamics are driven by the intricate interplay of competing mechanisms, such as social reinforcement and social weakening effects \citep{meng2025spreading}. These complex dynamics make disentangling latent interaction factors a significant challenge for traditional real-valued representation methods.
Moreover, symmetric modeling paradigms fail to capture the inherent asymmetry of social influence, where influence strengths are direction-dependent. For instance, the probability that a high-impact user triggers a response from a regular user is often significantly different from the reverse direction. Modeling these composite asymmetric relations using real-valued vectors usually requires additional antisymmetric matrices \citep{chang2019antisymmetricrnn}, dual-tower structures \citep{song2025boosting}, or complex feature-interaction modules \citep{he2025information}, which substantially increase model complexity and risk over-parameterization, thereby impairing generalization.

\begin{figure}
\centering
\includegraphics[width=0.55\textwidth]{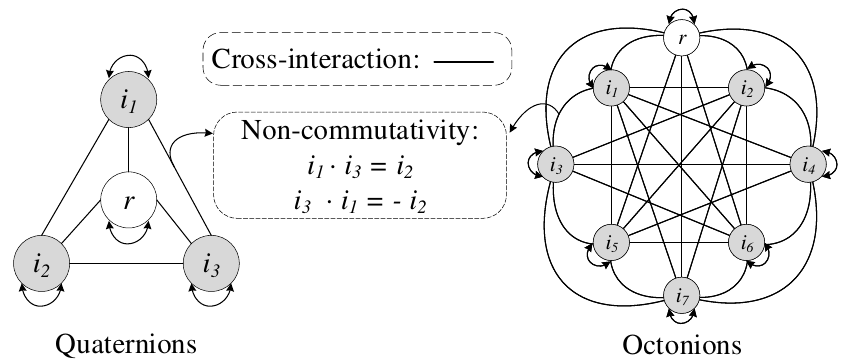}
\hfill
\caption{The operation properties of hypercomplex numbers (A case study of Quaternions and Octonions). $r$ denotes the real part, $i$ denotes the imaginary part.}
\label{fig:hypercomplex}
\end{figure}

Recently, frameworks based on hypercomplex algebra \citep{kantor1989hypercomplex,zhu2018quaternion} have achieved remarkable success in fields such as computer vision \citep{yu2024hypercomplex}, knowledge graphs \citep{li2020quaternion}, and recommendation systems \citep{li2025hypercomplex}, owing to their powerful high-dimensional representation capabilities. 
Formally, a hypercomplex number consists of one real component and multiple imaginary basis components, forming a unified algebraic entity that is not necessarily associative or commutative \citep{kantor1989hypercomplex}. As illustrated in Figure \ref{fig:hypercomplex}, the expanded component space and complex operation rules naturally correspond to cascade dynamics: multi-factor interactions correspond to the cross-interaction between various components, while the asymmetric influence is manifested as the non-commutativity of multiplication. These properties motivate us to explore hypercomplex algebra as a more expressive representation for cascade modeling.

In this paper, we propose CasHyper (\textbf{Cas}cade Prediction via \textbf{Hyper}complex Algebra), a hypercomplex based cascade prediction framework for modeling complex user interaction patterns. It innovatively integrates hypercomplex operations with Graph Neural Networks (GNNs) and Transformers to characterize multi-factor interactions among latent diffusion signals and asymmetric influence between users during information diffusion. In the network embedding stage, learnable user and cascade representations are organized into multiple hypercomplex components to model cross-component interactions among multiple latent factors. A static graph encoding channel aggregates structural information from the heterogeneous graph, while a Neural ODE-based dynamic channel models the continuous evolution of node states and edge weights. The non-commutative hypercomplex product provides order-dependent interaction representations, thereby giving the model the capacity to learn asymmetric diffusion patterns from data. In the cascade interaction stage, we design a hypercomplex Transformer that incorporates rotary positional transformations into the attention mechanism, allowing relative positional information to be encoded while preserving the norms of the rotated query and key representations. Finally, an interaction-aware prediction module and a linear projection module are adaptively integrated to estimate the probability of each candidate user participating next. Our main contributions are summarized as follows:
\begin{itemize}
\item We propose CasHyper, a unified hypercomplex framework for information cascade prediction. By organizing node embeddings into Cayley-Dickson components and introducing structured algebraic operations, CasHyper provides a flexible way to model cross-component interactions and direction-sensitive user influence.

\item We introduce a dual-channel network encoder that combines static heterogeneous graph aggregation with continuous dynamic encoding. The static channel captures user-user and user-cascade structural relations, while the Neural ODE-based channel jointly models the continuous evolution of node states and edge weights. 

\item We design a hypercomplex Transformer with rotary positional transformations for cascade sequence modeling. The proposed module introduces relative positional information into component-wise attention while preserving the norms of the rotated representations. 

\item We conduct extensive experiments on four real-world datasets, including comparisons with existing cascade prediction methods, controlled architectural variants, ablation studies, robustness and efficiency analyses, and an investigation of different Cayley-Dickson algebras. The results demonstrate the empirical effectiveness of CasHyper under the evaluated settings and reveal dataset-dependent behaviors of different hypercomplex structures.
\end{itemize}

The remainder of this paper is organized as follows. Section \ref{related work} reviews the related literature on cascade prediction, graph representation learning, and hypercomplex based applications. Section \ref{section:preliminaries} introduces the essential preliminaries, including the definitions of hypercomplex algebra and the formal statement of the cascade prediction task. In Section \ref{section:framework}, we detail the proposed CasHyper framework. Section \ref{section:experiments} presents the experimental setup, performance evaluation, and in-depth analysis. 
Finally, Section \ref{section:conclusion} concludes the paper and discusses broader theoretical and practical implications, as well as potential directions for future research.

\section{Related Work}
\label{related work}
\subsection{Cascade Prediction}
Early cascade prediction research primarily focused on the information propagation dynamics in social networks, aiming to simulate complex diffusion processes through mathematical models and feature analysis. \citet{shen2014modeling} pioneers a probabilistic model based on reinforced Poisson processes to simulate the dynamic arrival process of individual popularity, providing a foundational framework for information popularity prediction. \citet{bao2015modeling} further proposes a model based on self-exciting Hawkes processes, effectively capturing the dynamic characteristics of propagation by distinguishing the excitation strength of each retweeting event. Recognizing the close interplay between information diffusion and underlying network topology, \citet{farajtabar2017coevolve} employs temporal point processes to model their co-evolution, successfully simulating the dynamic trajectories of common propagation patterns.

Feature-driven models also play a significant role in early cascade prediction studies. \citet{cui2013cascading} constructs state matrices of information cascades based on node retweeting statuses, employing the Orthogonal Sparse Logistic Regression method to jointly optimize node selection and outbreak prediction. \citet{cheng2014can} analyzes temporal and structural features, revealing their critical role in predicting cascade scale. To combine the strengths of feature-driven and generative models, \citet{mishra2016feature} proposes a marked Hawkes self-exciting point process approach. By jointly modeling content virality, memory decay, and user influence, and training the prediction layer with historical cascade data, this method achieves more accurate popularity predictions.

With advancements in computational power, deep learning techniques, leveraging their robust feature extraction and ability to model complex patterns, have increasingly been applied to the cascade prediction. \citet{yang2019neural} employs attention mechanisms and convolutional networks, significantly enhancing the generalization capability of the approach. 
\citet{liao2019popularity} highlights the critical influence of tweet content on information diffusion, proposing a method that integrates hierarchical attention networks with recurrent neural networks (RNNs) to deeply fuse content features with temporal sequences for predicting the popularity of online content. 
Adopting a topic-oriented perspective, \citet{wang2021neural} incorporates thematic context into user representations through attention mechanisms, jointly modeling textual content and sequential information to more accurately predict retweeting behavior. 
\citet{wang2021dydiff} argues that user interests are typically influenced by dynamic social interactions, proposing a dynamic encoder to capture the temporal evolution of users' intrinsic interests over time, coupled with a dual-attention decoder to integrate content and propagation sequence information for estimating fusion likelihood. 
To better capture dynamic information during propagation, \citet{chen2019information} utilizes dynamic cascade subgraphs and Long Short-Term Memory (LSTM) to model the inherent dependencies between underlying topological structures and user retweeting behaviors, while \citet{wang2022casseqgcn} introduces node activation attributes to preserve the global structure of cascade graphs, thereby improving the integration of structural and temporal information. Going a step further in spatio-temporal modeling, \citet{sang2025dyhgtcr} constructs a dynamic heterogeneous graph based on repost and follow relationships, employing relative time encoding to extract unified spatio-temporal features that capture the intricate interplay between temporal evolution and social structures. Addressing the issue of mismatched macro- and micro-scale dynamics, \citet{yang2021full} proposes a reinforcement learning based framework that incorporates macro-level diffusion scale features into an RNN-based micro-level diffusion model, while integrating underlying social graph information to achieve joint prediction at both micro and macro levels. \citet{wang2023multiscale} focuses on the competitive and cooperative relationships between information, considering external factors such as content lifecycle, user preferences, and latent content expectations, and employs a minimal replacement theory to accomplish dual prediction tasks for popularity and retweeting behavior. \citet{liu2025gthnn} integrates graph transformers with sequential hypergraph neural networks to mitigate the over-smoothing and over-squashing issues common in traditional GNNs, effectively aggregating high-level social homogeneity and user dynamic preferences across global periods. Moving beyond traditional cascade observation, \citet{zhou2024modeling} enhances prediction accuracy across various time scales by capturing latent personalized retweeting patterns and temporal cascade characteristics. To tackle the out-of-distribution challenges, \citet{zhu2025ghidorah} leverages test-time training and invariant learning to mitigate spurious correlations and enhance model robustness against the inherent variability of social media data. Shifting the focus from individual-level to community-level dynamics, \citet{gao2025social} leverages large language models to enhance social context interpretation and mitigate the community cold-start problem. 

These methods demonstrate the importance of structural, temporal, and contextual information for cascade prediction. However, most of them still represent users and cascades in real-valued Euclidean spaces. In contrast, CasHyper explores a complementary representation direction by organizing node embeddings into hypercomplex components and using Cayley-Dickson products to construct direction-sensitive and component-coupled interaction representations.

\subsection{Graph Representation Learning}
Graph representation learning aims to encode nodes, edges, or graphs into low-dimensional vector spaces while preserving structural and semantic information. Early studies mainly relied on shallow network embedding methods. DeepWalk learns node embeddings by applying language-model-based objectives to truncated random-walk sequences \citep{perozzi2014deepwalk}. Node2vec further introduces biased random walks to balance breadth-first and depth-first exploration, enabling embeddings to capture both homophily and structural equivalence \citep{grover2016node2vec}. Although effective, these methods usually learn static embeddings and have limited ability to model complex nonlinear dependencies between node attributes and graph structures.

Graph neural networks have advanced graph representation learning through neural message passing and neighborhood aggregation. Graph Convolutional Networks jointly exploit node features and local graph topology by recursively aggregating neighborhood information \citep{kipf2017semi}. GraphSAGE extends this paradigm to inductive learning by learning aggregation functions for unseen nodes and evolving graphs \citep{hamilton2017inductive}. Graph Attention Networks further incorporate attention mechanisms to assign different importance weights to neighboring nodes \citep{velickovic2018graph}. These methods have established a general framework for learning expressive representations from graph-structured data. Graph autoencoders provide another important line of graph representation learning by encoding graph structures and node attributes into latent representations and reconstructing graph-related information through decoders. Variational Graph Autoencoders introduce probabilistic latent variables to model uncertainty in graph structures \citep{kipf2016variational}. To enhance the expressiveness of latent representations, \citet{xia2023multi} propose a multi-head variational graph autoencoder constrained by sum-product networks, which relaxes the mean-field assumption and improves fault tolerance.

Recent studies have further extended graph representation learning to more realistic scenarios involving incomplete, imbalanced, or signed graphs. \citet{xia2025incomplete} systematically review incomplete graph learning and highlight the importance of robust representation learning under missing attributes or incomplete topology. For attribute-missing graphs, \citet{xia2024attribute} propose Attribute Imputation Autoencoders to integrate observed attributes and structural information for more accurate attribute generation. Considering both attribute missingness and class imbalance, \citet{xia2025graphiam} develop GraphIAM, which combines graph contrastive learning with oversampling in pre-training and uses an adapter mechanism during fine-tuning. For signed graphs, \citet{gu2025mvsgae} propose a multi-head variational signed graph autoencoder to learn node representations from positive and negative subgraphs and infer both link existence and link signs.

In summary, graph representation learning has evolved from shallow network embedding to deep message-passing models, graph autoencoders, and robust learning frameworks for complex graph scenarios. Nevertheless, most existing methods still operate in real-valued vector spaces and mainly rely on aggregation, reconstruction, or contrastive objectives to capture graph dependencies. How to further enhance the expressive capacity of graph representations remains an important research problem.

\subsection{Hypercomplex Based Applications}
Hypercomplex algebras have demonstrated significant application value across various machine learning domains, including computer vision, natural language processing, and recommendation systems. \citet{parcollet2016quaternion} incorporates quaternion feature interactions into multilayer perceptrons (MLP) for language understanding tasks, effectively capturing complex feature dependencies in documents. \citet{trabelsi2017deep} further refines neural network modules by designing key components for complex-valued neural networks, including complex batch normalization and complex weight initialization strategies, achieving superior performance in audio processing tasks when applied to convolutional neural networks (CNNs) and LSTM networks. \citet{zhu2018quaternion} proposes a fully quaternion convolutional neural network, redesigning convolutional and fully connected layers in the quaternion domain, surpassing the performance of traditional real-valued CNNs in image classification and denoising tasks. Similarly, \citet{frants2023qcnn} develops an encoder-decoder architecture based on quaternion image representations, integrating quaternion per-pixel loss functions and normalization layers, significantly enhancing image dehazing performance. In recent years, hypercomplex algebras have also been introduced to recommendation systems. \citet{li2020quaternion} employs quaternion vectors to represent users and items, effectively capturing latent dependencies between entities. \citet{chen2021quaternion} proposes quaternion factorization machines and quaternion neural factorization machines, leveraging quaternion operations to model more complex feature interactions while substantially reducing the number of parameters, demonstrating higher efficiency compared to real-valued matrix multiplication. \citet{li2022hypercomplex} designs a hypercomplex graph convolution operator to model neighbor node interactions in hypercomplex spaces, thereby learning more robust representations of users and items. To enhance the expressiveness of entity representations, \citet{li2025hypercomplex} leverages multi-component hypercomplex embeddings and specialized graph neural networks to model the structural and semantic dependencies between users, items, and knowledge graph entities in a high-dimensional hypercomplex space.

Despite the notable success of hypercomplex algebra and advanced graph learning methods in these domains, their application to information diffusion modeling remains underexplored. Existing hypercomplex studies mainly focus on vision, language understanding, recommendation, or general graph representation learning, whereas cascade prediction requires jointly modeling social relations, historical user-cascade participation, temporal diffusion order, and asymmetric influence. CasHyper is designed for this setting by combining hypercomplex node representations, heterogeneous graph encoding, continuous dynamic edge modeling, and cascade sequence encoding within a unified prediction framework.

\begin{table}[t]
\centering
\caption{Summary of main notations.}
\label{tab:symbols}
\renewcommand{\arraystretch}{1.15}
\begin{tabular}{ll}
\toprule
\textbf{Symbol} & \textbf{Description} \\
\midrule
$G_s=(V_s,E_s)$ & Static directed social graph. \\
$V_s$, $E_s$ & Set of users, Set of directed social edges. \\
$\mathcal{C}$, $C_c$ & Set of cascades, The $c$-th information cascade. \\
$L_c$ & Length of cascade $C_c$. \\
$(u_i^c,t_i^c)$ & The $i$-th activated user and timestamp in $C_c$. \\
$G_b=(V_b,E_b)$ & User-cascade bipartite graph. \\
$G_h=(V_h,E_h)$ & Heterogeneous graph constructed from $G_s$ and $G_b$. \\
\midrule
$\mathcal{A}_m$ & The $m$-th algebra generated by the Cayley-Dickson construction. \\
$\mathbf{i}_k$ & The $k$-th hypercomplex basis unit. \\
$\Phi(\cdot)$ & Mapping from hypercomplex representation to real-valued coordinates. \\
$d$ & Total embedding dimension. \\
$d'$ & Dimension of each hypercomplex component, $d'=d/2^{m+1}$. \\
$\mathbf{X}_v$ & Initial learnable embedding of node $v$. \\
\midrule
$\mathbf{H}_s$ & Static node representations from neighborhood aggregation. \\
$\mathbf{X}^{(t)}$, $w_{e_{ij}}^{(t)}$ & Node states at continuous time $t$, Dynamic edge weight from $u_i$ to $u_j$ at time $t$. \\
$\mathbf{H}_d$ & Dynamic node representations from continuous encoding. \\
$\mathbf{H}_{hete}$ & Unified node representations after static-dynamic fusion. \\
\midrule
$\mathbf{H}_c$ & Cascade user representations retrieved from $\mathbf{H}_{hete}$. \\
$\mathbf{H}$ & Context-aware cascade representation. \\
$\mathbf{Q},\mathbf{K},\mathbf{V}$ & Query, key, and value tensors. \\
$\widetilde{\mathbf{Q}},\widetilde{\mathbf{K}}$ & Rotated query and key representations. \\
$\mathbf{H}'$ & Cascade representation with relative positional information. \\
$\mathbf{T}^c$ & Temporal interval representation of cascade $C_c$. \\
\midrule
$\widetilde{\mathbf{Y}}$, $\mathbf{Y}$ & Predicted probability distribution after softmax, Ground-truth next-user label. \\
$\mathcal{L}_{main}$, $\mathcal{L}_{cons}$ & Cross-entropy loss for next-user prediction, Consistency regularization loss. \\
\bottomrule
\end{tabular}
\end{table}

\section{Preliminaries}
\label{section:preliminaries}
Formally, a hypercomplex number can be expressed as,
\begin{equation}
    h_x = x_1 + x_2\mathbf{i}_2 + \cdots + x_n\mathbf{i}_n = x_1 + \sum_{k=2}^n x_k\mathbf{i}_k,
\end{equation}
where $x_1, x_2, \cdots, x_n$ are real coefficients, and $\mathbf{i}_2, \mathbf{i}_3, \cdots, \mathbf{i}_n$ are imaginary units obeying specific multiplication rules.
\subsection{Cayley-Dickson Construction}
The Cayley-Dickson construction \citep{kantor1989hypercomplex} is a recursive algebraic method for generating higher-dimensional algebraic structures from lower-dimensional ones. 
Starting from the real field $\mathcal{A}_0=\mathbb{R}$, each doubling step constructs a new algebra $\mathcal{A}_{m+1}$ from two copies of the previous algebra $\mathcal{A}_m$. Specifically, a new formal unit $e_m$ is introduced at the $m$-th doubling step, and each element in $\mathcal{A}_{m+1}$ can be written as,
\begin{equation}
x = a + b e_m, \quad a,b \in \mathcal{A}_m .
\end{equation}

For two elements $x=a+b e_m \in \mathcal{A}_{m+1}$ and $y=c+d e_m \in \mathcal{A}_{m+1}$, where $a,b,c,d \in \mathcal{A}_m$, the operations in $\mathcal{A}_{m+1}$ are defined recursively as follows,
\begin{itemize}
    \item The $addition$ of $x$ and $y$ is defined as:
            $ x \oplus_{m+1} y = (a \oplus_m c) + (b \oplus_m d)e_m .$ The $subtraction$ operation is defined in the same component-wise manner by replacing $\oplus$ with $\ominus$.
    \item The $conjugate$ of $x$ is defined recursively as: $\bar{x} = \bar{a} - b e_m .$ For the base algebra $\mathcal{A}_0=\mathbb{R}$, the conjugate of a real number $r \in \mathbb{R}$ is defined as $\bar{r}=r$.
    \item The $multiplication$ of $x$ and $y$ is defined recursively as: $x \otimes_{m+1} y = (a \otimes_m c \ominus_m \bar{d} \otimes_m b) + (d \otimes_m a \oplus_m b \otimes_m \bar{c})e_m.$ The order of factors and conjugates must be preserved, as multiplication is generally non-commutative.
    \item The $scalar multiplication$ of $x$ is defined as: $\lambda x = \lambda a + \lambda b e_m ,$ where $\lambda \in \mathbb{R}$ is a scalar.
    \item The $quadratic$ form associated with $x$ is defined recursively as: $||x||_{m+1} = ||a||_m + ||b||_m.$ For the base algebra $\mathcal{A}_0=\mathbb{R}$, it is defined as $||r||_0=r^2$ for $r\in\mathbb{R}$.
\end{itemize}

\subsection{Problem Definition}
\textbf{Definition 1 (Social Graph).} Let $G_s=(V_s,E_s)$ denote a static, directed, and unweighted social graph, where $V_s$ is the set of users and $E_s \subseteq V_s \times V_s$ is the set of directed edges representing social relations between users. A directed edge $(u_i,u_j)\in E_s$ indicates that there exists a social relation from user $u_i$ to user $u_j$.

\textbf{Definition 2 (Information Cascade).} The propagation process of a message $c$ in a social network is defined as an information cascade. Different from an unordered set, a cascade is represented as a time-ordered sequence $C_c = \left\langle (u_1^c,t_1^c), (u_2^c,t_2^c), \ldots, (u_{L_c}^c,t_{L_c}^c) \right\rangle$, where $u_i^c \in V_s$ denotes the $i$-th user participating in cascade $C_c$, $t_i^c$ denotes the corresponding participation time, and $L_c$ denotes the length of the cascade. The sequence is ordered chronologically, satisfying $t_1^c \leq t_2^c \leq \cdots \leq t_{L_c}^c$. User participation may correspond to actions such as reposting, commenting, or retweeting.

\textbf{Definition 3 (User-Cascade Bipartite Graph).} Given a collection of historical cascades $\mathcal{C}=\{C_1,C_2,\ldots,C_{|\mathcal{C}|}\}$, we construct a user-cascade bipartite graph $G_b=(V_b,E_b)$ to model historical user-cascade participation relations. The node set $V_b$ consists of two disjoint subsets,
$$V_b = V_\mathcal{U} \cup V_\mathcal{C}, \quad V_\mathcal{U} \cap V_\mathcal{C} = \emptyset,$$
where $V_\mathcal{U} \subseteq V_s$ denotes the set of users appearing in the historical cascades, and $V_\mathcal{C}$ denotes the set of cascade nodes, with each node $V_c$ corresponding to a historical cascade $C_c$. The edge set is defined as, $E_b \subseteq V_\mathcal{U} \times V_\mathcal{C}$. Specifically, an edge $(u_i,c_c)\in E_b$ exists if and only if user $u_i$ participates in cascade $C_c$. In this work, $G_b$ is treated as a static and unweighted bipartite graph constructed from historical cascade records.

\textbf{Problem Formulation.} Given the social graph $G_s$ and the historical user-cascade bipartite graph $G_b$, the task of information cascade prediction is to forecast the future evolution of an ongoing cascade. Formally, for a newly observed partial cascade $C_p = \left\langle (u_1^p,t_1^p), (u_2^p,t_2^p), \ldots, (u_{L_p}^p,t_{L_p}^p) \right\rangle$, where $u_i^p\in V_s$ and $t_1^p \leq t_2^p \leq \cdots \leq t_{L_p}^p$, the objective is to predict the next participant $u_{L_p+1}^p$. The candidate user set is defined as $U_{\mathrm{cand}} = V_s \setminus \{u_1^p,u_2^p,\ldots,u_{L_p}^p\}.$ We aim to learn a prediction model that estimates the conditional probability of each candidate user participating next,
$$\hat{p}(u) = P\left(u=u_{L_p+1}^p \mid C_p,G_s,G_b\right), \quad \forall u\in U_{\mathrm{cand}}.$$
The predicted next participant is then obtained by
$\hat{u}_{L_p+1} = {\arg\max}_{u\in U_{\mathrm{cand}}} \hat{p}(u).$ For clarity, we summarize the main notations used throughout this paper in Table~\ref{tab:symbols}.

\section{Proposed Framework}
\label{section:framework}
\begin{figure*}
\centering
\includegraphics[width=1\textwidth]{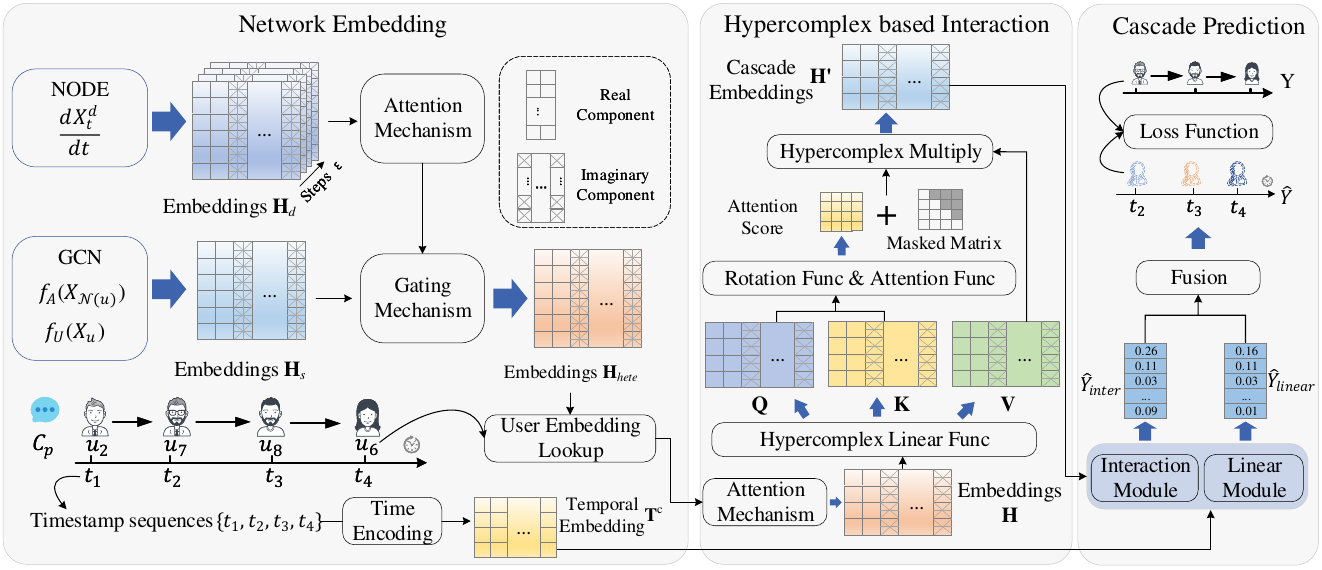}
\caption{The framework of CasHyper.}
\label{fig:framework}
\end{figure*}

This section introduces the design of CasHyper. The overall framework of CasHyper, as illustrated in Figure \ref{fig:framework}, comprises three components: Network Embedding, Hypercomplex based Interaction, and Cascade Prediction. 

\subsection{Input Construction}
\begin{figure}
\centering
\includegraphics[width=0.75\textwidth]{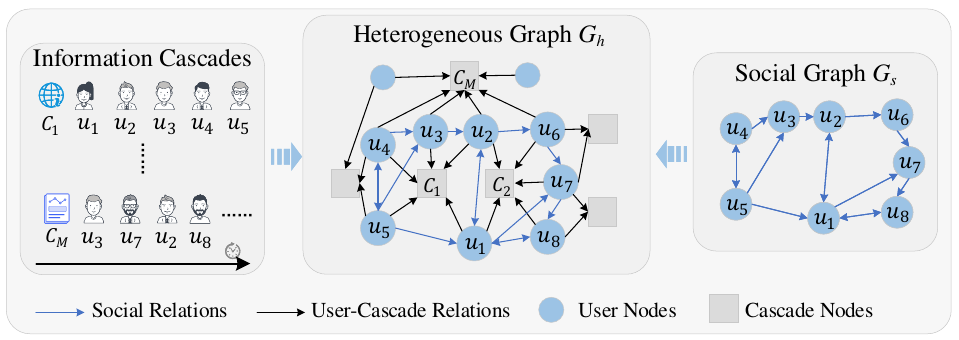}
\hfill
\caption{The input construction of CasHyper.}
\label{fig:input_construct}
\end{figure}

To jointly model user–cascade interaction patterns during information diffusion, we construct a heterogeneous graph. While many contemporary methods \citep{wang2022cascade,wang2024information} combine the social graph ($G_s$), the bipartite graph ($G_b$), and the fine-grained cascade diffusion graphs, the latter are often unavailable in practice.
Therefore, our model only relies on the social graph $G_s=(V_s,E_s)$ and the historical user-cascade bipartite graph $G_b=(V_b,E_b)$, where $V_b=V_\mathcal{U}\cup V_\mathcal{C}$, $V_\mathcal{U}\cap V_\mathcal{C}=\emptyset$, $V_\mathcal{U}\subseteq V_s$, and $V_\mathcal{C}$ denotes the set of historical cascade nodes. Specifically, the constructed heterogeneous graph is denoted as $G_h=(V_h,E_h)$, where $V_h$ and $E_h$ denote the node set and edge set, respectively. The node set of $G_h$ is defined as $V_h = V_s \cup V_\mathcal{C}$, the edge set is defined as $E_h = E_s \cup E_b$. Accordingly, $G_h$ contains two types of nodes, namely users and cascades, and two types of edges, namely user-user edges and user-cascade edges. This construction enables the model to jointly capture social relations among users and historical participation relations between users and cascades. The construction process is illustrated in the input module shown in Figure \ref{fig:input_construct}.

\subsection{Network Embedding}
Since explicit node attributes are unavailable, we initialize both user and cascade nodes with learnable ID embeddings. For each node $v\in V_h$, we assign a learnable embedding vector $\mathbf{x}_v \in \mathbb{R}^{d}$. These embeddings are randomly initialized and jointly optimized during training.
To construct node embeddings in a higher-order hypercomplex space, we utilize the Cayley-Dickson construction to encode nodes as elements of the algebra $\mathcal{A}_{m+1}$. Specifically, node features $\mathbf{X}$ are partitioned into $2^{m+1}$ components, yielding a structured representation $\mathbf{X} \in \mathbb{R}^{|V_h| \times 2^{m+1} \times d/2^{m+1}}$, where each component corresponds to a distinct part of the $\mathcal{A}_{m+1}$ algebra.
For user node $u$ and cascade node $c_c$, their feature representations are defined as $\mathbf{X}_u$ and $\mathbf{X}_{c}$, respectively. Taking the $\mathbf{X}_u$ as an example,
\begin{equation}
    \textbf{X}_{u} = \textbf{a}_u + \textbf{b}_u e_m = \textbf{X}_{u}^{1} + \sum\nolimits_{k=2}^{2^{m+1}} \textbf{X}_{u}^{k}\textbf{i}_k,
\end{equation}
where $\mathbf{a}_u,\mathbf{b}_u\in\mathcal{A}_m$ are two subalgebra components of $\mathbf{X}_u$, $e_m$ is the newly introduced unit in the current Cayley-Dickson doubling step. $\mathbf{X}_{u}^{k} \in \mathbb{R}^{d/2^{m+1}}$ is the real-valued representation for component $k$. Notably, while the hypercomplex algebra encompasses multiple components, the total embedding dimension is fixed at $d$ (e.g., $d=64$ and $d'=d/2^{m+1}$) to prevent an increase in model parameters.

\subsubsection{Static Neighborhood Aggregation}
\label{subsec:Static Neighborhood Aggregation}
Given a node $u$ and its incoming neighbors $\mathcal{N}_u^{\mathrm{in}}$, we update the static representation of the central node by summing the hypercomplex representations of its neighbors, as follows,
\begin{equation}
\begin{aligned}
    \mathbf{h}_{s,u}^{l+1} &= \text{AGG}(\mathbf{h}_{s,v}^l |v \in \mathcal{N}_u^{\mathrm{in}}) = \sum\nolimits_{v \in \mathcal{N}_u^{\mathrm{in}}} \frac{1}{p_{uv}} \mathbf{h}_{s,v}^l \\
    &= \sum\nolimits_{v \in \mathcal{N}_u^{\mathrm{in}}} \frac{1}{p_{uv}} \mathbf{a}_{v}^l + \sum\nolimits_{v \in \mathcal{N}_u^{\mathrm{in}}} \frac{1}{p_{uv}} \mathbf{b}_{v}^l e_m \\
    &= \sum\nolimits_{k=1}^{2^{m+1}}\sum\nolimits_{v \in \mathcal{N}_u^{\mathrm{in}}}\frac{1}{p_{uv}}\mathbf{h}_{s,v}^{l,k}\mathbf{i}_k,
\label{equ:hypercomplex_agg}
\end{aligned}
\end{equation}
where $l$ is the layer number, $\mathbf{h}_{s, v}^0 = \mathbf{X}_v$ is node $v$'s initial embedding, and $\mathbf{h}_{s,v}^{l,k}\in\mathbb{R}^{d'}$ denotes the $k$-th real-valued component. $\mathbf{a}_v^l, \mathbf{b}_v^l \in \mathcal{A}_m$ are the subalgebras of $\mathbf{h}_{s, v}^l \in \mathcal{A}_{m+1}$. 
$p_{uv} = \sqrt{|\mathcal{N}_u^{\mathrm{in}}||\mathcal{N}_v^{\mathrm{out}}|}$ denotes the direction-aware normalization constant, where $\mathcal{N}_u^{\mathrm{in}}$ and $\mathcal{N}_v^{\mathrm{out}}$ denote the overall directed neighborhoods used for aggregation. Since different types of neighbors are aggregated jointly in our heterogeneous graph, the neighborhood sets contain all relevant node and edge types rather than being separated by relation type.

Since hypercomplex addition is performed component-wise over the basis $\mathbf{i}_k$, this aggregation is algebraically identical to element-wise feature aggregation in standard GCNs \citep{kipf2017semi}. Let $\Phi: \mathcal{A}_{m+1} \rightarrow \mathbb{R}^d$ be the vector space isomorphism that maps a hypercomplex representation to its real coordinate vector. 
For a single node $u$, the static representation after aggregation can be written as,
\begin{equation}
\mathbf{h}_{s,u}^{l+1}=f_U\big(\Phi(\mathbf{h}_{s,u}^{l}), f_A\big({\Phi(\mathbf{h}_{s,v}^{l}) | v \in \mathcal{N}_u^{\mathrm{in}}}\big)\big),
\label{equ}
\end{equation}
where $f_A$ is the neighborhood aggregation function and $f_U$ is the update function that integrates the current representation of node $u$ with the aggregated neighbor information.
Equivalently, for all nodes in the heterogeneous graph, the component-wise aggregation can be written in matrix form as,
\begin{equation}
\nonumber
\mathbf{H}_{s}^{l+1}=
\tilde{\mathbf{A}}_h
\mathbf{H}_{s}^{l},
\label{equ}
\end{equation}
where $\tilde{\mathbf{A}}_h$ is the normalized adjacency matrix of $G_h$, and $\mathbf{H}_{s}^{l}\in\mathbb{R}^{|V_h|\times 2^{m+1}\times d'}$ denotes the component-wise hypercomplex node representations at the $l$-th layer. The multiplication by $\tilde{\mathbf{A}}_h$ is performed along the node dimension and is shared across all hypercomplex components. Therefore, the hypercomplex structure is preserved by maintaining the component partitioning, while the aggregation itself can be efficiently implemented using standard real-valued tensor operations.

\subsubsection{Continuous Asymmetric Encoding} 
The continuous dynamic characteristics of the propagation process are also critical. Motivated by \citet{wang2024information}, we introduce a continuous dynamic encoding channel based on Neural Ordinary Differential Equations (Neural ODEs) on top of the traditional GNN encoders. The core idea is to treat both the node states $\mathbf{X}^{(t)}$ and the edge weights $w^{(t)}$ as variables that evolve continuously over time $t$. We formulate a coupled system of ODEs, wherein the dynamics of edge weights depend on the node states, and in turn, the evolution of node states is governed by the dynamically changing graph structure. 
For an edge from user $u_i$ to user $u_j$, the edge weight $w_{e_{ij}}^{(t)}$ is defined as a real-valued function of the hypercomplex product between the two endpoint node states,
\begin{equation}
\begin{aligned}
    \frac{d w_{{e}_{ij}}^{(t)}}{dt} &= f(\mathbf{X}_{u_i}^{(t)} \otimes_{m+1} \mathbf{X}_{u_j}^{(t)})
    = \sigma \big(\mathrm{MLP}(\Phi(\mathbf{X}_{u_i}^{(t)} \otimes_{m+1} \mathbf{X}_{u_j}^{(t)}))\big),
\end{aligned}
\label{equ:dhe}
\end{equation}
where $\Phi(\cdot)$ maps a hypercomplex-valued representation to its real-coordinate vector, $\mathrm{MLP}(\cdot)$ denotes a multi-layer perceptron, and $\sigma(\cdot)$ is the sigmoid function. Therefore, $f: \mathcal{A}_{m+1}\rightarrow \mathbb{R}{\geq 0}$ maps the hypercomplex-valued product to a nonnegative real scalar. Using the Cayley-Dickson multiplication rule, the hypercomplex product in Eq.~\eqref{equ:dhe} can be expanded as,
\begin{equation}
\nonumber
\begin{aligned}
    &\mathbf{X}_{u_i}^{(t)} \otimes_{m+1} \mathbf{X}_{u_j}^{(t)} = (\mathbf{a}_{u_i}^{(t)} + \mathbf{b}_{u_i}^{(t)}e_m) \otimes_{m+1} (\mathbf{a}_{u_j}^{(t)} + \mathbf{b}_{u_j}^{(t)}e_m) \\
    &= (\mathbf{a}_{u_i}^{(t)} \otimes_m \mathbf{a}_{u_j}^{(t)}  \ominus_m \overline{\mathbf{b}_{u_j}^{(t)}} \otimes_m \mathbf{b}_{u_i}^{(t)}) + (\mathbf{b}_{u_j}^{(t)} \otimes_m \mathbf{a}_{u_i}^{(t)} \oplus_m \mathbf{b}_{u_i}^{(t)} \otimes_m \overline{\mathbf{a}_{u_j}^{(t)}})e_m,
\end{aligned}
\end{equation}
where $\mathbf{a}_{u_i}^{(t)}, \mathbf{b}_{u_i}^{(t)}, \mathbf{a}_{u_j}^{(t)}, \mathbf{b}_{u_j}^{(t)} \in \mathcal{A}_m$ are subalgebras components of $\mathbf{X}_{u_i}^{(t)}$ and $\mathbf{X}_{u_j}^{(t)}$. $w_{e_{ij}}^{(t)} \in \mathbb{R}$ is the asymmetric edge weight from $u_i$ to $u_j$ at time step $t$. 
A key advantage of this formulation is its inherent ability to model the asymmetric influence between nodes. The intuition stems directly from the non-commutative nature of the hypercomplex product $\otimes_{m+1}$.

\paragraph{Asymmetry-aware Representation.} For two distinct nodes $u_i$ and $u_j$, the continuous asymmetric encoding based on the Cayley-Dickson multiplication can produce order-dependent interaction representations, 
\begin{equation} \nonumber
 \mathbf{X}_{u_i}^{(t)} \otimes_{m+1} \mathbf{X}_{u_j}^{(t)} \neq \mathbf{X}_{u_j}^{(t)} \otimes_{m+1} \mathbf{X}_{u_i}^{(t)}. 
\end{equation}
Therefore, the model is structurally capable of distinguishing the influence from $u_i$ to $u_j$ and that from $u_j$ to $u_i$, providing an algebraic inductive bias for modeling asymmetric influence in cascade prediction. A detailed discussion of this property is provided in Appendix~\ref{appendix:asymmetric influence}.

At any time $t$, the dynamic edge weights $w^{(t)}$, obtained by integrating Eq. \eqref{equ:dhe}, are used to update the graph structure, denoted as $G_h^t =\{V_h, E_h, w^{(t)} \}$. The dynamic evolution of the node embeddings $\mathbf{X}^{(t)}$ is then governed by an ODE parameterized by a GCN encoder $\Psi$,
\begin{equation} 
\frac{d \mathbf{X}^{(t)}}{dt} = \Psi(G_h^t, \mathbf{X}^{(t)}) - \mathbf{X}^{(t)} + \mathbf{X}^{(0)}. 
\label{equ:dXt_condensed} 
\end{equation}
This formulation defines the instantaneous rate of change of the node states as a combination of three terms: (1) The GCN encoder, $\Psi$, propagates information on the time-varying graph $G_h^t$ based on the current state $\mathbf{X}^{(t)}$. (2) A decay term, $-\mathbf{X}(t)$, acts as a self-correction mechanism that prevents states from growing unboundedly. (3) An initial state, $\mathbf{X}^{(0)}$, ensures that the evolution remains anchored to the initial state.
We employ a numerical ODE solver, Runge-Kutta, to integrate the coupled system defined by Eq. \eqref{equ:dhe} and Eq. \eqref{equ:dXt_condensed} over $\epsilon$ time intervals, yielding a sequence of node embeddings $\{\mathbf{X}^{(0)}, \mathbf{X}^{(1)}, \cdots, \mathbf{X}^{(\epsilon)} \}$.

To obtain a single dynamic representation $\mathbf{H}_{d} \in \mathbb{R}^{|V_h| \times 2^{m+1} \times d'}$, we apply an attention-based temporal aggregation mechanism. Specifically, we compute a node-wise attention vector $\boldsymbol{\alpha}_i \in \mathbb{R}^{|V_h|}$ for each time step $i$,
\begin{equation}
    \boldsymbol{\alpha}_i = \frac{\text{exp}(\Phi(\textbf{X}^{(i)})\textbf{W}_{\alpha} \mathbf{a}) }{\sum\nolimits_{j=0}^{\epsilon} \text{exp} (\Phi(\textbf{X}^{(j)})\textbf{W}_{\alpha} \mathbf{b})},\ \ \ \ \mathbf{H}_{d} = \sum\nolimits_{i \in \epsilon} \boldsymbol{\alpha}_i\textbf{X}^{(i)},
\label{equ:aggratation}
\end{equation}
where $\textbf{a} \in \mathbb{R}^d$ and $\textbf{W}_{\alpha} \in \mathbb{R}^{d \times d}$ are trainable parameters.

After obtaining the static node embeddings $\mathbf{H}_{s}$ and dynamic node embeddings $\mathbf{H}_{d}$, we adopt an adaptive gating mechanism to integrate them into a unified representation $\mathbf{H}_{hete} \in \mathbb{R}^{|V_h| \times 2^{m+1} \times d'}$. Specifically, we first compute two node-wise gating scores,
\begin{equation}
[\boldsymbol{\beta}_{s}, \boldsymbol{\beta}_{d}] = \big[\operatorname{softmax}\big(\Phi(\mathbf{H}_{s})\mathbf{W}_{s}\mathbf{b}_{s}\big), \operatorname{softmax}\big(\Phi(\mathbf{H}_{d})\mathbf{W}_{d}\mathbf{b}_{d}\big)\big], \quad \mathbf{H}_{hete} = \boldsymbol{\beta}_{s} \mathbf{H}_{s} + \boldsymbol{\beta}_{d} \mathbf{H}_{d}, 
\label{equ:fusing-Hd-Hs}
\end{equation} 
where $\mathbf{W}_s, \mathbf{W}_d \in \mathbb{R}^{d \times d}$ and $\mathbf{b}_s, \mathbf{b}_d \in \mathbb{R}^d$ are trainable parameters, $\boldsymbol{\beta}_{s}, \boldsymbol{\beta}_{d} \in \mathbb{R}^{|V_h|}$ are node-wise gating vectors. The second dimension of $\mathbf{H}_{hete}$ corresponds to the $2^{m+1}$ components, which can be interpreted as the different parts of the $(m+1)$-th hypercomplex algebra.

\subsection{Hypercomplex based Interaction}
After obtaining the hypercomplex representation $\mathbf{H}_{hete}$, we model the interactions within the cascade sequence through a two-stage process: a context-driven attention stage followed by explicit relative position encoding.

In the following, we clarify the tensor-level hypercomplex projection used in the subsequent equations. For the representation $\mathbf{R} \in \mathbb{R}^{L \times 2^{m+1} \times d'}$ and a learnable hypercomplex projection matrix $\mathbf{W}_R \in \mathbb{R}^{2^{m+1} \times d'}$, the notation $\mathbf{R} \otimes_{m+1} \mathbf{W}_R$ denotes a tensor-level hypercomplex projection along the feature dimension,
\begin{equation}
\nonumber
(\mathbf{R} \otimes_{m+1} \mathbf{W}_R)_{l} = \mathbf{R}_{l} \otimes_{m+1} \mathbf{W}_R, \quad l=1,\ldots,L, 
\end{equation}
where $\otimes_{m+1}$ on the right-hand side denotes the Cayley-Dickson multiplication between $2^{m+1}$ hypercomplex components. Therefore, $\mathbf{R} \otimes_{m+1} \mathbf{W}_R \in \mathbb{R}^{L \times 2^{m+1} \times d'}$. This operation is performed independently for each cascade position and preserves the hypercomplex-component dimension. 
The use of Cayley-Dickson multiplication also introduces structured cross-component interactions. Different hypercomplex components are not transformed independently. Instead, they are coupled according to the algebraic multiplication rules of the corresponding Cayley-Dickson algebra. This provides an explicit algebraic inductive bias for modeling interactions among different latent factors of cascade states and user representations.

\subsubsection{Context Driven Relationships}
Given a cascade $C_c$ with a maximum length of $L_c$, we retrieve user representations $\textbf{H}_{c} \in \mathbb{R}^{L_c \times 2^{m+1} \times d'}$ from $\textbf{H}_{hete}$ and apply masked attention mechanism to obtain a context-aware representation $\textbf{H} \in \mathbb{R}^{L_c \times 2^{m+1} \times d'}$,
\begin{equation}
    \textbf{H} = \text{softmax}_j\big(\frac{\Phi(\textbf{H}_{c} \otimes_{m+1} \textbf{W}_{Q_c}) \cdot \Phi(\textbf{H}_{c} \otimes_{m+1} \textbf{W}_{K_c})^\top}{\sqrt{d}} + \textbf{M}_c \big) \textbf{H}_{c},
\label{eq:self_attention}
\end{equation}
where $\textbf{W}_{Q_c}, \textbf{W}_{K_c} \in \mathbb{R}^{2^{m+1} \times d'}$ are learnable hypercomplex projection parameters, $\textbf{M}_c \in \mathbb{R}^{L_c \times L_c}$ is a causal mask matrix containing 0 and $-\infty$ to shield future information, and $\mathrm{softmax}_j$ denotes row-wise normalization. The mapping function $\Phi(\cdot)$ converts the hypercomplex-valued tensors $\mathbb{R}^{L_c \times 2^{m+1} \times d'}$ into real-valued representations $\mathbb{R}^{L_c \times d}$. Therefore, the score matrix inside the softmax operation is real-valued and forms an attention matrix in $\mathbb{R}^{L_c \times L_c}$. This real-valued attention matrix is multiplied with $\mathbf{H}_{c}$ along the cascade length dimension, with the same attention weights applied component-wise to all $2^{m+1}$ hypercomplex components. While this stage restricts each node to attend only to its preceding nodes through autoregressive masking, it does not explicitly model the precise relative positions between nodes.

\subsubsection{Relative Positional Relationships} To inject explicit positional information, we introduce the hypercomplex Transformer that encodes the order and distance of nodes into the interaction computation. Specifically, we first map the cascade user representations $\mathbf{H}$ into the query $\textbf{Q} \in \mathbb{R}^{L_c  \times 2^{m+1} \times d'}$, key $\textbf{K} \in \mathbb{R}^{L_c  \times 2^{m+1} \times d'}$, and value $\textbf{V} \in \mathbb{R}^{L_c  \times 2^{m+1} \times d'}$ tensors through tensor-level hypercomplex projections,
\begin{equation}
    \textbf{Q} = \textbf{H} \otimes_{m+1} \mathbf{W}_Q,\quad  \textbf{K} = \textbf{H} \otimes_{m+1} \mathbf{W}_K,\quad  \textbf{V} = \textbf{H} \otimes_{m+1} \mathbf{W}_V,
\label{equ:GenerateQ-K-V}
\end{equation}
where $\mathbf{W}_Q, \mathbf{W}_K, \mathbf{W}_V \in \mathbb{R}^{2^{m+1} \times d'}$ are learnable hypercomplex projection parameters. Subsequently, the frequency $\bm{\omega}^Q, \bm{\omega}^K \in \mathbb{R}^{L_c  \times 2^{m+1} \times d'}$ and phase angle $\bm{\theta}^Q, \bm{\theta}^K \in \mathbb{R}^{L_c  \times 2^{m+1} \times d'}$ for rotation are generated based on the $\textbf{Q}$ and $\textbf{K}$ to integrate the node contextual information during the rotation process,
\begin{equation}
\begin{split}
    \bm{\omega}^Q = \text{ReLU}(\textbf{Q} \otimes_{m+1} \mathbf{W}_{\omega}), \quad
    \bm{\theta}^Q = \text{Tanh}(\textbf{Q} \otimes_{m+1} \mathbf{W}_{\theta}), \\
    \bm{\omega}^K = \text{ReLU}(\textbf{K} \otimes_{m+1} \mathbf{W}_{\omega}), \quad
    \bm{\theta}^K = \text{Tanh}(\textbf{K} \otimes_{m+1} \mathbf{W}_{\theta}),
\end{split}
\label{equ:Generate-frequency-phase}
\end{equation}
where $\mathbf{W}_{\omega}, \mathbf{W}_{\theta} \in \mathbb{R}^{2^{m+1} \times d'}$ are learnable transformation parameters. The rotation transformation can be formally expressed as,
\begin{equation}
% \begin{aligned}
    \widetilde{\textbf{Q}} = ||_{k=1}^{2^{m+1}}\textbf{Q}_{k} \cdot e^{\textbf{i}(2\pi\bm{\omega}_{k}^{Q}pos\ +\ \bm{\theta}_{k}^Q)}_{k}, \quad 
    \widetilde{\textbf{K}} = ||_{k=1}^{2^{m+1}}\textbf{K}_{k} \cdot e^{\textbf{i}(2\pi\bm{\omega}_{k}^Kpos\ +\ \bm{\theta}_{k}^K)}_{k},
% \end{aligned}
\label{equ:rotation}
\end{equation}
where $||$ is concatenate operation, $\mathbf{Q}_k, \mathbf{K}_k, \bm{\omega}_k^Q, \bm{\omega}_k^K, \bm{\theta}_k^Q, \bm{\theta}_k^K,  \in \mathbb{R}^{L_c \times d'}$ represent the $k$-th real-valued components of their corresponding hypercomplex representations.
$\widetilde{\textbf{Q}}, \ \widetilde{\textbf{K}} \in \mathbb{R}^{2^{m+1} \times L_c \times d'}$ are rotated matrices under the component-first notation, $pos = [1, 2, \cdots, L_c]$ denotes absolute position of nodes, $e^{\textbf{i}(2\pi\bm{\omega_k}^Qpos\ +\ \bm{\theta_k}^Q)} = \text{cos}(2\pi\bm{\omega}_{k}^Q pos\  +\  \bm{\theta}_{k}^Q)\  +\  \textbf{i}\ \text{sin}(2\pi\bm{\omega}_{k}^Q pos + \bm{\theta}_{k}^Q)$ represents the rotation parameters constructed based on Euler's formula, where $\textbf{i}$ is a unit hypercomplex number satisfying a unit norm for its imaginary part.
In our representation, the hypercomplex representation is implemented as a set of real-valued component vectors associated with different hypercomplex basis units. The $k$-th hypercomplex component vector $\mathbf{v}_k \in \mathbb{R}^{d'}$ denotes the real-valued coefficient vector corresponding to the $k$-th basis component. 
The hypercomplex rotational transformation follows the standard unit-modulus rotation property, which helps inject positional information without changing the magnitude of each component representation. We summarize this property as follows, with a brief explanation provided in Appendix~\ref{appendix:norm preservation}.
\begin{property}[\textnormal{Norm Preservation}]
For any hypercomplex component vector $\mathbf{v}_k$, its L2-norm remains invariant under the rotation $\mathcal{R}$:
$||\mathcal{R}(\mathbf{v}_k)||_2 = ||\mathbf{v}_k||_2$.
\label{property:norm preservation}
\end{property}

Thus, positional information is injected without distorting semantic magnitudes. The cascade representations $\mathbf{H}' \in \mathbb{R}^{L_c \times 2^{m+1} \times d'}$ incorporating relative positional relationships can be expressed as follows,
\begin{equation} 
\mathbf{H}'_{:,k,:} = f_{\sigma} \left( \mathbf{S}_k \mathbf{V}_{:,k,:} + \mathbf{H}_{:,k,:} \right), \quad \mathbf{S}_k = \operatorname{softmax}_{j} \left( \frac{ \langle \widetilde{\mathbf{Q}}_{k}, \widetilde{\mathbf{K}}_{k} \rangle }{\sqrt{d'}} + \mathbf{M}_{r,k} \right), \quad k=1,\ldots,2^{m+1}, 
\label{eq:rotation_att_score} 
 \end{equation} 
 where $\mathbf{V}_{:,k,:}, \mathbf{H}_{:,k,:}, \widetilde{\mathbf{Q}}_{k}, \widetilde{\mathbf{K}}_{k} \in \mathbb{R}^{L_c \times d'}$ denote the $k$-th components. $\mathbf{S}_k \in \mathbb{R}^{L_c \times L_c}$ is the real-valued attention matrix for the $k$-th hypercomplex component, and $\operatorname{softmax}_{j}$ denotes row-wise normalization over the key-position dimension, $\langle \  \cdot \ , \ \cdot \ \rangle $ denotes the real part of the Hermitian inner product. The mask $\mathbf{M}_{r,k} \in \mathbb{R}^{L_c \times L_c}$ is the relative-position attention mask matrix containing 0 and $-\infty$ for the $k$-th component. $f_{\sigma}$ is a linear mapping with an activation function. After aggregating all components along the hypercomplex component dimension, we obtain $\mathbf{H}' \in \mathbb{R}^{L_c \times 2^{m+1} \times d'}$. The core of this mechanism is that by rotating the query and key vectors ($\widetilde{\textbf{Q}}$ and $\widetilde{\textbf{K}}$) according to their absolute positions, the conjugate product depends on relative position. This key property can be formalized as follows,
\begin{property}[\textnormal{Relative Position Dependence}]
When the query and key rotary parameters are shared or approximately aligned, i.e., $\bm{\omega}^Q \approx \bm{\omega}^K$ and $\bm{\theta}^Q \approx \bm{\theta}^K$, the real part of the Hermitian inner product between the rotated query vector $\widetilde{\mathbf{Q}}_j$ at position $j$ and the rotated key vector $\widetilde{\mathbf{K}}_l$ at position $l$ is a function of their semantic information and relative position $(j-l)$.
\label{property:relative position dependence}
\end{property}

The explanation, detailed in Appendix~\ref{appendix:relative position dependence}, is based on a regularization condition that the query and key rotary parameters are shared or approximately aligned, i.e., $\bm{\omega}^Q \approx \bm{\omega}^K$ and $\bm{\theta}^Q \approx \bm{\theta}^K$. To explicitly encourage the model to satisfy this condition during optimization, we introduce a consistency regularization term based on the Mean Squared Error (MSE). This loss penalizes the discrepancy between the rotational parameters of the query ($\bm{\omega}^Q, \bm{\theta}^Q$) and the key ($\bm{\omega}^K, \bm{\theta}^K$),
\begin{equation}
\mathcal{L}_{\text{cons}} = \frac{1}{L_c} \left( ||\bm{\omega}^Q - \bm{\omega}^K||_F^2 + ||\bm{\theta}^Q - \bm{\theta}^K||_F^2 \right),
\label{equ:consistency loss}
\end{equation}
where $||\cdot||_F^2$ denotes the squared Frobenius norm.

\subsection{Cascade Prediction}
The time intervals between user actions are a critical indicator of a cascade's momentum \citep{cao2017deephawkes}. Shorter intervals often signal rising popularity, while longer ones may suggest waning influence. This motivates our approach to explicitly encode temporal features.

Given a cascade $C_c$ with a time series $\{t_1^c, t_2^c, \cdots, t_{L_c}^c\}$, we first compute the sequence of time intervals $\bm{\delta}^c=\{0, \delta_2^c, \delta_3^c, \cdots, \delta_{L_c}^c\}$, $\delta_1^c=0$ and $\delta_{i+1}^c=t_{i+1}^c-t_i^c$ for $i \ge 1$. To effectively preserve the discrete distributional characteristics of temporal patterns while reducing the complexity of continuous time data, we predefine a set of discrete time interval bins $\{(0, T_1],  (T_1, T_2], \cdots, (T_{k-1}, T_k]\}$, where $T_k$ represents the maximum timestamp of observed cascades. For each time interval $\delta_i^c$, we identify its corresponding bin and represent it as a $k$-dimensional one-hot encoded vector (with 1 indicating the activated bin and 0 otherwise), collectively forming the temporal interval matrix $\textbf{T}_{\delta}^c \in \{0,1\}^{L_c \times k}$.
A neural projection produces an enhanced temporal representation $\textbf{T}^c \in \mathbb{R}^{L_c \times d_t}$,
\begin{equation}
    \textbf{T}^c = \sigma(\textbf{W}_t\textbf{T}_{\delta}^c + \textbf{b}_t),
\label{equ:cal-time-intervals}
\end{equation}
where $\textbf{W}_t \in \mathbb{R}^{k \times d_t}$ and $\textbf{b}_t \in \mathbb{R}^{d_t}$ are learnable parameters. The bias $\mathbf{b}_t$ is broadcast along the cascade length dimension.
With both the cascade representation $\textbf{H}'$ and the enhanced temporal representation $\textbf{T}^c$, we propose two modules for predicting the next user to be activated.

\textbf{Interaction-Aware Module.}
This module explicitly models fine-grained user-cascade affinity. We compute scores by interacting the evolving cascade state $\textbf{H}' \in \mathbb{R}^{L_c \times 2^{m+1} \times d'}$ with user hypercomplex embeddings $\textbf{H}_{hete}^u \in \mathbb{R}^{|V_s| \times 2^{m+1} \times d'}$,
\begin{equation}
    \hat{\textbf{Y}}_{inter} = \Phi(\textbf{H}') \cdot \Phi(\textbf{H}_{hete}^u)^{\top} + \textbf{M}_I,
\label{equ:inter}
\end{equation}
where $\hat{\textbf{Y}}_{inter} \in \mathbb{R}^{L_c \times |V_s|}$ represents the specific appeal of the cascade to users at each time step, and $\textbf{M}_I \in \mathbb{R}^{L_c \times |V_s|}$ is a mask matrix used to exclude already activated nodes.

\textbf{Linear Projection Module.} For computational efficiency, we concatenate $\mathbf{H}'$ with $\mathbf{T}^c$ to form a temporally-enriched embedding $(\Phi(\textbf{H}')||\textbf{T}^c) \in \mathbb{R}^{L_c \times (d + d_t)}$, mapped directly to the target space,
\begin{equation}
    \hat{\textbf{Y}}_{linear} = \textbf{W}_l(\Phi(\mathbf{H}')||\textbf{T}^c)^{\top} + \textbf{M}_L.
\label{equ:linear}
\end{equation}
Finally, we employ a fusion layer to integrate the predictive outputs of both modules,
\begin{equation}
    \hat{\textbf{Y}} = \mathbf{\gamma} \hat{\textbf{Y}}_{inter} + (1-\mathbf{\gamma}) \hat{\textbf{Y}}_{linear}, \ \ \ \mathbf{\gamma} = \sigma(\hat{\mathbf{Y}}_{inter} \cdot \mathbf{w}_{\gamma}),
\label{equ:final_logits}
\end{equation}
where $\sigma$ is the Sigmoid function, $\mathbf{w}_{\gamma} \in \mathbb{R}^{|V_s|}$ is a learnable parameter, $\mathbf{\gamma} \in \mathbb{R}^{L_c}$ is a position-wise weight coefficient.

\subsection{Model Optimization}
We formulate the main objective as the cross-entropy loss for next-user prediction, summed over all cascades and positions,
\begin{equation}
    \mathcal{L}_\text{main} = - \sum_{c=1}^{|C|}\sum_{i=2}^{L_c}\sum_{j=1}^{|V_s|}\textbf{Y}_{ij}^c\text{log}(\widetilde{\textbf{Y}}_{ij}^c), \quad \widetilde{\mathbf{Y}}_{ij}^c = \frac{\exp(\hat{\mathbf{Y}}_{ij}^c)} {\sum_{r=1}^{|V_s|}\exp(\hat{\mathbf{Y}}_{ir}^c)}, 
\label{equ:final_score}
\end{equation}
where $\mathbf{Y}_{ij}^c=1$ indicates that, in cascade $C_c$, user $u_j$ is the $i$-th activated user based on the previously observed users $\{u_1^c,\ldots,u_{i-1}^c\}$, and $\mathbf{Y}_{ij}^c=0$ otherwise. The predicted probability distribution over candidate users is obtained by applying the softmax function along the candidate-user dimension. The summation starts from $i=2$ because the first activated user is the source node, for which no historical activation sequence is available for next-user prediction.

The model is trained end-to-end by minimizing a joint loss function that integrates the primary prediction objective with the consistency regularization term. The total loss $\mathcal{L}$ is defined as,
\begin{equation}
\mathcal{L} = \mathcal{L}_{\text{main}} + \lambda_{\text{cons}}\mathcal{L}_{\text{cons}},
\label{equ:total_loss}
\end{equation}
where $\lambda_{\text{cons}} \geq 0$ is a weighting coefficient that controls the contribution of the consistency regularizer. The objective is to optimize predictive accuracy while encouraging alignment of the query and key rotary parameters through the consistency regularization term. We use $\lambda_{\mathrm{cons}}=1.0$ as the default setting through experiment analysis.

\begin{algorithm}[t]
\caption{Pseudocode of CasHyper}
\label{alg:cashyper}
\begin{algorithmic}[1]
\REQUIRE Social graph $G_s=(V_s,E_s)$; historical cascades $\mathcal{C}$; embedding dimension $d$; Cayley-Dickson order $m$; ODE integration intervals $\epsilon$; regularization coefficient $\lambda_{\mathrm{cons}}$.
\ENSURE Predicted next user $\hat{u}_{L_p+1}$ and optimized model parameters $\Theta$.

\STATE Construct the user-cascade bipartite graph $G_b=(V_b,E_b)$ from historical cascades $\mathcal{C}$.
\STATE Construct the heterogeneous graph $G_h=(V_h,E_h)$ by combining $G_s$ and $G_b$.
\STATE Initialize learnable node embeddings and partition them into $2^{m+1}$ hypercomplex components.

\STATE Perform static neighborhood aggregation on $G_h$ to obtain $\mathbf{H}_s$. \quad $\triangleright$ Eq.~\eqref{equ:hypercomplex_agg}

\FOR{$t=0$ to $\epsilon-1$}
    \STATE Compute dynamic asymmetric edge weights $w_{e_{ij}}^{(t)}$.  \quad $\triangleright$ Eq.~\eqref{equ:dhe}
\ENDFOR
\STATE Update the time-varying heterogeneous graph $G_h^t =\{V_h, E_h, w^{(t)} \}$.
    \STATE Compute the dynamic evolution of the node embeddings $\mathbf{X}^{(t)}$. \quad $\triangleright$ Eq.~\eqref{equ:dXt_condensed}
\STATE Aggregate continuous node states to obtain $\mathbf{H}_d$. \quad $\triangleright$ Eq.~\eqref{equ:aggratation}
\STATE Compute $\mathbf{H}_{hete}$ through fusing $\mathbf{H}_s$ and $\mathbf{H}_d$. \quad $\triangleright$ Eq.~\eqref{equ:fusing-Hd-Hs}

\FOR{each cascade $C_c$ in the training batch}
    \STATE Retrieve cascade user representations $\mathbf{H}_c$ from $\mathbf{H}_{hete}$.
    \STATE Apply context-driven masked attention to obtain $\mathbf{H}$. \quad $\triangleright$ Eq.~\eqref{eq:self_attention} 
    \STATE Generate hypercomplex query, key, and value tensors $\mathbf{Q}$, $\mathbf{K}$, and $\mathbf{V}$. \quad $\triangleright$ Eq.~\eqref{equ:GenerateQ-K-V}
    \STATE Generate rotary frequency $\bm{\omega}^Q, \bm{\omega}^K$ and phase $\bm{\theta}^Q, \bm{\theta}^K$. \quad $\triangleright$ Eq.~\eqref{equ:Generate-frequency-phase}  
    \STATE Apply rotary transformation to obtain $\widetilde{\mathbf{Q}}$ and $\widetilde{\mathbf{K}}$. \quad $\triangleright$ Eq.~\eqref{equ:rotation} 
    \STATE Compute cascade representations $\mathbf{H}'$. \quad $\triangleright$ Eq.~\eqref{eq:rotation_att_score}
    \STATE Encode time intervals into temporal representation $\mathbf{T}^c$. \quad $\triangleright$ Eq.~\eqref{equ:cal-time-intervals}
    \STATE Compute prediction scores using the interaction-aware module $\hat{\textbf{Y}}_{inter}$ and the linear projection module $\hat{\textbf{Y}}_{linear}$. \quad $\triangleright$ Eq.~\eqref{equ:inter} and Eq.~\eqref{equ:linear}
    \STATE Fuse the two prediction scores to obtain the final logits $\hat{\mathbf{Y}}$. \quad $\triangleright$ Eq.~\eqref{equ:final_logits}
    \STATE Apply softmax over candidate users to obtain $\widetilde{\mathbf{Y}}$. \quad $\triangleright$ Eq.~\eqref{equ:final_score}
\ENDFOR

\STATE Compute the total loss $\mathcal{L}$. \quad $\triangleright$ Eq.~\eqref{equ:total_loss}
\STATE Optimize the parameters by back-propagation.

\RETURN Optimized model parameters $\Theta$.
\end{algorithmic}
\end{algorithm}

\subsection{Model Complexity}
We analyze the theoretical complexity of CasHyper within the scope of an iteration. Let $d$ denote the total embedding dimension, $n=2^{m+1}$ denote the number of components in the hypercomplex number, and $d'=d/2^{m+1}$ denote the dimension of each component.

In the network embedding stage, the static neighborhood aggregation based on an $L$-layer GCN incurs a complexity of $\mathcal{O}\left(L(|E_h|d+|V_h|d^2)\right)$, where $|V_h|$ and $|E_h|$ denote the numbers of nodes and edges in the heterogeneous graph, respectively.
In the continuous asymmetric encoding stage, the complexity comprises dynamic edge-weight computation and node-state evolution. For each edge, computing the dynamic edge weight between two $n$-component representations has complexity $\mathcal{O}(n^2d')$. Therefore, the cost of dynamic edge-weight computation over all edges is $\mathcal{O}\left(|E_h|(n^2d')\right)$. The node-state evolution parameterized by the L-layer GCN encoder $\Psi$ has complexity $\mathcal{O}\left(L(|E_h|d+|V_h|d^2)\right)$.
Given $\epsilon$ numerical integration steps in the ODE solver, the overall complexity of the continuous asymmetric encoding stage is $\mathcal{O}\left(\epsilon\left[|E_h (n^2d')+L(|E_h|d+|V_h|d^2)\right]\right)$.
In the hypercomplex interaction stage, the computational cost is mainly dominated by hypercomplex projections, Euler transformations, and rotary attention. The hypercomplex projection and Euler transformation introduce costs of $\mathcal{O}(L_cn^2d')$ and $\mathcal{O}(L_cnd')$, respectively, while the rotary attention computation costs $\mathcal{O}(L_c^2nd')$. Therefore, the overall complexity of this stage is $\mathcal{O}\left(L_cn^2d' + L_cnd' + L_c^2nd'\right)$, where $L_c$ denotes the cascade sequence length. The quadratic term with respect to $L_c$ becomes the main bottleneck when modeling long cascade sequences.

In the cascade prediction stage, CasHyper computes prediction scores over the full user set. Specifically, the interaction-aware module requires $\mathcal{O}(L_c|V_s|d)$, and the linear projection module requires $\mathcal{O}(L_c|V_s|(d+d_t))$.
The final softmax normalization over candidate users further incurs $\mathcal{O}(L_c|V_s|)$ complexity. In our implementation, we use a full softmax over all users in $V_s$, with activated users masked out during prediction. This exact scoring strategy ensures a fair comparison with the reported evaluation protocol, but its cost grows linearly with the number of users and can become expensive for very large social networks. Incorporating approximate candidate generation or sampled softmax is a promising direction for improving scalability in future work.

\section{Experiments}
\label{section:experiments}
\subsection{Experimental Setups}
\textbf{Datasets.} We evaluate our model on four widely used real-world cascade datasets: Android \citep{sankar2020inf}, Christianity \citep{sankar2020inf}, Twitter \citep{hodas2014simple}, and Memetracker \citep{leskovec2009meme}. These datasets cover diverse platforms, including forum discussions, social media, and online news diffusion. For consistency, we use the term node to denote the activated entity in each dataset. In user-centric datasets, nodes correspond to users, whereas in source-centric datasets such as Memetracker, nodes correspond to URLs or web sources.
\begin{itemize}
\item \textbf{Android} and \textbf{Christianity} datasets, sourced from the Stack Exchange network of community Q\&A websites, represent knowledge-sharing social networks in technical and humanities domains. In these datasets, the social network is formed through various interpersonal interaction channels, such as posting questions, providing answers, and leaving comments. Different from repost-based platforms, where explicit diffusion paths can be directly observed, Stack Exchange records topic-centered knowledge interactions among users. Following this setting, we construct cascades as chronologically ordered sequences of posts associated with the same topical tag. Such cascades should be interpreted as topic-driven discussion cascades, which characterize how information and attention around a specific topic evolve over time within the community. Therefore, they serve as a proxy for information diffusion in knowledge-sharing communities, rather than as explicit repost or forwarding cascades. This construction is suitable for the next-user prediction task because it models the temporal order of user participation under a shared topical information context.
\item \textbf{Twitter} focuses on the diffusion of information markers in the form of URLs. By integrating the follower-followee social graph with tweet metadata, this dataset tracks how users are exposed to and subsequently post specific URLs. In this work, we define a URL reposting event as the case where a user posts a URL that has previously appeared in their social feed. Accordingly, cascades in this dataset represent exposure-based URL diffusion processes.
\item \textbf{Memetracker} monitors the migration of frequent quotes and phrases (memes) across the global web. Unlike user-centric social network datasets, Memetracker describes the spread of textual information across web sources. In this dataset, individual URLs are treated as activated source nodes for modeling purposes, and the diffusion of a meme across different URLs over time constitutes an information cascade. Therefore, the prediction task on Memetracker is more precisely interpreted as next-source prediction, i.e., predicting which URL/source will mention the meme next. This setting is still compatible with our general cascade prediction framework, where an activated node can denote either a user node or an information-source node, depending on the dataset.
\end{itemize}

\begin{table}
  \centering
  \begin{threeparttable}
  \caption{Statistics of the datasets.}
    \begin{tabular}{ccccc}
    \toprule
    Datasets & Android & Christianity & Twitter & Memetracker \\
    \midrule
    \# Users & 2,927 & 2,897 & 12,627 & 4,709 \\
    \# Links & 48,573 & 35,624 & 309,631 & 31,448 \\
    \# Cascades & 678   & 583   & 3,461 & 12,634 \\
    \#Avg. Length & 39.41 & 23.20  & 30.39    & 15.65 \\
    \bottomrule
    \end{tabular}%
    \label{tab:statistics}%
    \end{threeparttable}

\end{table}%

\textbf{Implementation Details.} 
For each dataset, we randomly split 80\% of the cascades for training, 10\% for validation, and the remaining 10\% for testing. All baseline models are evaluated on the same splits for fair comparison. To avoid information leakage, the user--cascade bipartite graph $G_b$ and the heterogeneous graph $G_h$ are constructed using only the training cascades. For each prediction position, only the previously observed prefix of the cascade is used as input, while the target node is kept unseen and used only for evaluation. Consistent with prior studies \citep{qiao2023rotdiff,wang2024information}, we treat cascade prediction as an information retrieval task and adopt Hits@k (hits score on top k) and MAP@k (mean average precision on top k) as evaluation metrics. In this paper, k is set to 10 and 50. For model parameters, the node representation dimension $d$ is set to 64, the number of graph neural network layers for learning node representation vectors is set to 2, and the time step $\epsilon$ for modeling continuous node dynamics in NODE is set to 3. The number of time interval bins for partitioning is set to $10^4$, and the dimension of time information embedding $d_t$ is set to 8. 
% For baseline models, we retained their publicly reported parameter settings. 
For baseline models, we re-implement and re-execute them using their official codebases on the same dataset splits as CasHyper to ensure a fair comparison.
All model parameters are updated using the Adam optimizer, with parameters $\beta1$ and $\beta2$ set to 0.90 and 0.98, respectively. The learning rate is set to 0.001.

\subsection{Baselines}
To evaluate the performance of the CasHyper model, we select ten state-of-the-art baselines for comparison. The following provides a concise overview of each baseline model. 
\begin{itemize}
    \item \textbf{NDM} \citet{yang2019neural} constructs a model based on two heuristic assumptions, integrating attention mechanisms and convolutional networks to model cascades, achieving better fitting of diffusion data. 
    \item\textbf{FOREST} \citet{yang2021full} incorporates macroscopic diffusion scale information into a GRU-based microscopic diffusion model via reinforcement learning, enabling full-scale cascade prediction. 
    \item\textbf{DyHGCN} \citet{yuan2020dyhgcn} proposes a dynamic heterogeneous graph network to encode social and dynamic diffusion graph structures, capturing users’ dynamic preferences. 
    \item\textbf{CE-GCN} \citet{wang2022cascade} explicitly integrates collaborative patterns between users and cascades into user embeddings through heterogeneous graphs and message-passing modules. 
    \item\textbf{MS-HGAT} \citet{sun2022ms} introduces a memory-enhanced sequential hypergraph attention network to model users’ global dependencies and dynamic preferences comprehensively. 
    \item\textbf{DisenIDP} \citet{cheng2023enhancing} employs lightweight hypergraph convolution to capture users’ potential intents from multiple perspectives, combined with attention-based encoders to extract long-term and short-term cascade influences. 
    \item\textbf{RotDiff} \citet{qiao2023rotdiff} maps each user to a Lorentzian vector, leveraging rotational transformations in hyperbolic space to model local diffusion dependencies within a given cascade. 
    \item\textbf{GODEN} \citet{wang2024information} utilizes neural ordinary differential equations to describe the co-evolutionary dynamics of nodes and edges, inferring users’ dynamic preferences based on the solutions. 
    \item\textbf{MetaCas} \citet{cheng2025information} designs two meta-knowledge-aware topological-temporal modules to extract user correlations during information diffusion through dynamic parameter generation. 
    \item\textbf{PMRCA} \citet{he2025pattern} employs structural contrastive learning anchored in node homogeneity and clustering contrastive learning driven by cognitive clusters to characterize the information diffusion.
\end{itemize}

While Large Language Models (LLMs) have demonstrated strong capabilities in semantic analysis, they are inapplicable to our experimental setting due to the absence of textual information in the evaluated datasets. Consequently, our baseline selection focuses on state-of-the-art topology-based and sequence-based diffusion models.

\begin{table}[!htbp]\setlength{\tabcolsep}{1.0mm}
  \centering
  \newcommand{\pms}[1]{\pm \text{\scriptsize #1}}
  \caption{Experimental results on Hits score over four datasets (\%). CasHyper uses the quaternion algebra $\mathbb{H}$ as the default algebraic type in the main experiments.}
    \begin{tabular}{l|cc|cc|cc|cc}
    \toprule
    \multirow{2}[5]{*}{Model} & \multicolumn{2}{c|}{Android}  & \multicolumn{2}{c|}{Christianity} & \multicolumn{2}{c|}{Twitter}   & \multicolumn{2}{c}{Memetracker} \\
\cmidrule{2-9}          & @10    & \multicolumn{1}{c|}{@50}    & @10    & \multicolumn{1}{c|}{@50}     & @10    & \multicolumn{1}{c|}{@50}    & @10    & \multicolumn{1}{c}{@50} \\
    \midrule
    NDM\textsuperscript{2019}   & $3.33\pms{0.17}$  & $9.29\pms{0.31}$   & $15.46\pms{0.46}$ & $35.35\pms{0.29}$  & $25.95\pms{0.29}$  & $37.26\pms{0.62}$  & $20.42\pms{0.08}$ & $36.01\pms{0.25}$ \\
    FOREST\textsuperscript{2021}  & $7.00\pms{1.31}$     & $16.29\pms{1.53}$  & $24.94\pms{0.53}$ & $47.82\pms{0.74}$  & $29.77\pms{1.34}$ & $44.18\pms{1.52}$  &$ 29.73\pms{1.13}$ & $48.04\pms{1.38}$ \\
    DyHGCN\textsuperscript{2020}  & $8.36\pms{0.57}$  & $19.08\pms{0.61}$   & $26.22\pms{0.51}$ & $50.64\pms{0.58}$ & $31.48\pms{0.93}$ & $50.43\pms{0.98}$  & $29.78\pms{0.43}$ & $48.79\pms{0.32}$ \\
    CE-GCN\textsuperscript{2022}  & $7.06\pms{0.35}$  & $16.38\pms{0.33}$  & $29.24\pms{0.83}$ & $50.45\pms{0.85}$  & $29.94\pms{0.71}$ & $49.21\pms{0.83}$  & $37.81\pms{0.95}$ & $56.73\pms{0.91}$ \\
    MS-HGAT\textsuperscript{2022}  & $9.08\pms{0.25}$  & $19.79\pms{0.52}$  & $28.99\pms{0.85}$ & $47.34\pms{1.13}$ & $32.66\pms{0.71}$ & $47.86\pms{0.68}$  & $28.11\pms{0.35}$ & $49.95\pms{0.18}$ \\
    DisenIDP\textsuperscript{2023} & $11.34\pms{0.76}$ & $20.67\pms{0.28}$  & $31.03\pms{0.36}$ & $52.90\pms{0.69}$  & $30.46\pms{0.48}$ & $50.46\pms{0.52}$ & $30.97\pms{0.81}$ & $52.54\pms{0.93}$ \\
    RotDiff\textsuperscript{2023} & $11.27\pms{0.17}$ & \underline{$23.31\pms{1.07}$}  & $30.80\pms{0.79}$  & $51.34\pms{0.14}$ & $34.18\pms{1.34}$ & $51.58\pms{1.91}$ & $29.72\pms{0.42}$ & $50.35\pms{0.37}$ \\
    GODEN\textsuperscript{2024} & \underline{$12.11\pms{0.21}$} & $21.43\pms{0.42}$  & $31.70\pms{0.48}$  & $52.23\pms{1.04}$ & $35.33\pms{0.89}$ & $54.22\pms{0.94}$ & $42.26\pms{0.36}$ & $61.98\pms{0.73}$ \\
    MetaCas\textsuperscript{2025} & $9.31\pms{0.33}$  & $21.23\pms{0.27}$  & $28.10\pms{0.35}$  & $52.43\pms{0.75}$ & \underline{$35.82\pms{0.47}$} & \underline{$55.00\pms{0.63}$} & $31.19\pms{0.73}$ & $50.50\pms{0.82}$ \\
    PMRCA\textsuperscript{2025} & $10.63\pms{0.55}$ & $17.79\pms{0.73}$  & \underline{$34.62\pms{0.29}$} & \underline{$54.09\pms{0.31}$} & $33.68\pms{0.22}$ & $46.52\pms{0.27}$ & \underline{$46.56\pms{0.78}$} & \underline{$64.74\pms{0.71}$} \\
    CasHyper & $\mathbf{16.07\pms{0.13}}$ & $\mathbf{33.39\pms{0.42}}$  & $\mathbf{37.50\pms{1.65}}$ & $\mathbf{64.06\pms{1.42}}$  & $\mathbf{40.20\pms{0.12}}$  & $\mathbf{59.20\pms{0.11}}$  & $\mathbf{48.19\pms{0.61}}$ & $\mathbf{67.29\pms{0.57}}$ \\
    \midrule
    Improve.(\%) & 32.70 & 43.24 & 8.32  & 18.43  & 12.23 & 7.64 & 3.50  & 3.94 \\
    \bottomrule
    \end{tabular}%
    \label{tab:results_hits}%
\end{table}%

\begin{table}[!htbp]\setlength{\tabcolsep}{1.0mm}
  \centering
  \newcommand{\pms}[1]{\pm \text{\scriptsize #1}}
  \caption{Experimental results on MAP score over four datasets (\%). CasHyper uses the quaternion algebra $\mathbb{H}$ as the default algebraic type in the main experiments.}
    \begin{tabular}{l|cc|cc|cc|cc}
    \toprule
    \multirow{2}[5]{*}{Model} & \multicolumn{2}{c|}{Android}  & \multicolumn{2}{c|}{Christianity} & \multicolumn{2}{c|}{Twitter}   & \multicolumn{2}{c}{Memetracker} \\
\cmidrule{2-9}          & @10    & \multicolumn{1}{c|}{@50}    & @10    & \multicolumn{1}{c|}{@50}   & @10    & \multicolumn{1}{c|}{@50}    & @10    & \multicolumn{1}{c}{@50}  \\
    \midrule
    NDM\textsuperscript{2019}   & $1.33\pms{0.09}$  & $1.58\pms{0.08}$  & $5.88\pms{0.25}$  & $6.74\pms{0.22}$  & $18.23\pms{}0.20$  & $18.73\pms{}0.22$  & $10.37\pms{0.04}$ & $11.08\pms{0.05}$ \\
    FOREST\textsuperscript{2021}  & $4.05\pms{0.94}$  & $4.45\pms{0.98}$ & $11.90\pms{0.37}$  & $12.90\pms{0.41}$  & $20.50\pms{0.56}$  & $21.15\pms{0.61}$  & $16.40\pms{0.91}$  & $17.25\pms{0.93}$ \\
    DyHGCN\textsuperscript{2020}  & $4.55\pms{0.13}$  & $5.00\pms{0.14}$  & $12.11\pms{0.19}$ & $13.21\pms{0.23}$ & $19.63\pms{0.43}$ & $20.47\pms{0.58}$  & $16.31\pms{0.47}$ & $17.20\pms{0.51}$ \\
    CE-GCN\textsuperscript{2022}  & $4.15\pms{0.26}$  & $4.52\pms{0.28}$  & $18.35\pms{0.23}$ & $19.29\pms{0.25}$ & $17.94\pms{0.75}$ & $18.82\pms{0.73}$  & $22.51\pms{0.18}$ & $23.41\pms{0.25}$ \\
    MS-HGAT\textsuperscript{2022}  & $5.86\pms{0.16}$  & $6.33\pms{0.32}$  & $17.54\pms{0.41}$ & $18.38\pms{0.37}$ & $20.43\pms{0.68}$ & $21.13\pms{0.65}$ & $15.29\pms{0.13}$ & $16.28\pms{0.11}$ \\
    DisenIDP\textsuperscript{2023} & $6.86\pms{0.53}$  & $7.25\pms{0.51}$  & $18.31\pms{0.35}$ & $19.32\pms{0.32}$ & $15.26\pms{0.36}$ & $16.19\pms{0.38}$ & $16.21\pms{0.21}$ & $17.22\pms{0.32}$ \\
    RotDiff\textsuperscript{2023}  & $6.80\pms{0.23}$   & $7.31\pms{0.27}$ & $19.03\pms{0.52}$ & $19.97\pms{0.60}$ & $22.36\pms{1.08}$ & $23.16\pms{1.11}$ & $15.47\pms{0.19}$ & $16.41\pms{0.19}$ \\
    GODEN\textsuperscript{2024} & \underline{$7.71\pms{0.15}$}  & \underline{$8.12\pms{0.14}$} & $20.12\pms{0.56}$ & \underline{$21.55\pms{0.53}$} & $22.15\pms{0.33}$ & $23.01\pms{0.31}$ & $24.02\pms{0.46}$ & $24.97\pms{0.51}$ \\
    MetaCas\textsuperscript{2025} & $4.80\pms{0.09}$   & $5.31\pms{0.09}$ & $13.52\pms{0.21}$ & $14.62\pms{0.16}$ & \underline{$22.90\pms{0.93}$}  & \underline{$23.79\pms{0.95}$}  & $16.92\pms{0.53}$ & $17.82\pms{0.58}$ \\
    PMRCA\textsuperscript{2025} & $5.83\pms{0.44}$   & $6.15\pms{0.41}$ & \underline{$20.14\pms{0.16}$} & $21.09\pms{0.18}$ & $22.72\pms{0.36}$  & $23.31\pms{0.32}$  & \underline{$25.10\pms{0.23}$} & \underline{$26.89\pms{0.18}$} \\
    CasHyper & $\textbf{8.59}\pms{0.21}$  & $\textbf{9.36}\pms{0.18}$  & $\textbf{21.95}\pms{0.88}$ & $\textbf{23.21}\pms{0.83}$  & $\textbf{25.11}\pms{0.15}$ & $\textbf{26.00}\pms{0.22}$  & $\textbf{26.87}\pms{0.38}$ & $\textbf{27.80}\pms{0.34}$ \\
    \midrule
    Improve.(\%) & 11.41  & 15.27  & 8.99  & 7.70  & 9.65 & 9.29   & 7.05  & 3.38 \\
    \bottomrule
    \end{tabular}%
    \label{tab:results_map}%
\end{table}%

\subsection{Experimental Results}
The overall performance comparison between CasHyper and various state-of-the-art baselines is summarized in Table \ref{tab:results_hits} and Table \ref{tab:results_map}, the best results are highlighted in bold, while underlined numbers denote the second-best performance. The last row reports the relative improvement of CasHyper over the best baseline for each metric. As observed, our proposed model consistently surpasses the real-valued baseline models across all datasets with significant margins. For instance, on the Android and Christianity datasets, CasHyper achieves a remarkable improvement of up to $43.24\%$ and $18.43\%$ in $Hits@50$, respectively. These empirical results validate the effectiveness of hypercomplex algebra for modeling asymmetric and cross-component interactions between nodes.

A key advantage of CasHyper lies in its ability to model asymmetric influence and cross-component interactions. Although geometric models like RotDiff employ hyperbolic space to capture the hierarchical nature of social networks, they often struggle with non-hierarchical, diverse interaction patterns. The non-commutative nature of hypercomplex multiplication allows CasHyper to more naturally distinguish between the directional influences of different users. Furthermore, our hypercomplex Transformer, integrated with norm-preserving rotary transformations, ensures that fine-grained interaction patterns and relative positional relationships are preserved without the information loss common in standard sequential models. 

Interestingly, the experimental results reveal that the improvement in MAP metrics is generally less pronounced than the gains in Hits. We attribute this disparity to the ranking noise inherent in real-world diffusion data. $Hits@k$ primarily measures the model's ability to identify the correct nodes (recall), whereas $MAP@k$ is highly sensitive to the exact order of activations. In practice, the recorded timestamps often suffer from coarse granularity or are influenced by exogenous factors such as user login patterns and content discovery times. Since a later timestamp does not necessarily reflect a weaker user preference or a different diffusion priority, the scope for precise ranking optimization is naturally constrained. Nevertheless, the substantial boost in Hits indicates that CasHyper provides a superior ability to capture the reach of a cascade within the social topology, even when the exact temporal sequence is stochastic.

\begin{table}[!htb]\setlength{\tabcolsep}{0.8mm} 
  \centering
  \newcommand{\pms}[1]{\pm \text{\scriptsize #1}}
  \begin{threeparttable}
  \caption{Results of different hypercomplex algebras on Hits Metric (\%).}
    \begin{tabular}{l|cc|cc|cc|cc}
    \toprule
    \multirow{2}[5]{*}{Type} & \multicolumn{2}{c|}{Android}  & \multicolumn{2}{c|}{Christianity} & \multicolumn{2}{c|}{Twitter}   & \multicolumn{2}{c}{Memetracker} \\
\cmidrule{2-9}          & @10    & \multicolumn{1}{c|}{@50}    & @10    & \multicolumn{1}{c|}{@50}    & @10    & \multicolumn{1}{c|}{@50}    & @10    & \multicolumn{1}{c}{@50} \\
    \midrule
    $\mathbb{C}$ & $15.68\pms{0.27}$ & $31.99\pms{0.53}$  & $37.11\pms{0.92}$ & $62.50\pms{1.26}$  & $40.10\pms{0.15}$  & $58.72\pms{0.25}$ & $38.98\pms{1.13}$ & $60.46\pms{0.95}$ \\
    $\mathbb{H}$ & $16.07\pms{0.13}$ & $\textbf{33.39}\pms{0.42}$  & $37.50\pms{1.65}$  & $64.06\pms{1.42}$ & $40.20\pms{0.12}$  & $59.20\pms{0.11}$  & $\textbf{48.19}\pms{0.61}$ & $\textbf{67.29}\pms{0.57}$ \\
    $\mathbb{O}$ & $\textbf{16.53}\pms{0.35}$ & $32.37\pms{0.38}$ & $\textbf{38.06}\pms{1.33}$ & $\textbf{64.50}\pms{1.13}$ & $\textbf{40.45}\pms{0.12}$ & $\textbf{59.75}\pms{0.28}$ & $46.59\pms{0.79}$ & $65.77\pms{0.43}$ \\
    $\mathbb{S}$ & $15.76\pms{0.21}$ & $32.68\pms{0.33}$  & $36.60\pms{1.21}$  & $64.06\pms{1.19}$ & $40.19\pms{0.17}$ & $59.68\pms{0.21}$ & $43.36\pms{0.93}$ & $65.14\pms{0.54}$ \\
    \bottomrule
    \end{tabular}%
  \label{tab: hyper_types_hits}%
  \end{threeparttable}
\end{table}%

\begin{table}[!htb]\setlength{\tabcolsep}{0.8mm} 
  \centering
  \newcommand{\pms}[1]{\pm \text{\scriptsize #1}}
  \begin{threeparttable}
  \caption{Results of different hypercomplex algebras on MAP Metric (\%).}
    \begin{tabular}{l|cc|cc|cc|cc}
    \toprule
    \multirow{2}[5]{*}{Type} & \multicolumn{2}{c|}{Android}  & \multicolumn{2}{c|}{Christianity} & \multicolumn{2}{c|}{Twitter}   & \multicolumn{2}{c}{Memetracker} \\
\cmidrule{2-9}          & @10    & \multicolumn{1}{c|}{@50}    & @10    & \multicolumn{1}{c|}{@50}    & @10    & \multicolumn{1}{c|}{@50}   & @10    & \multicolumn{1}{c}{@50} \\
    \midrule
    $\mathbb{C}$  & $8.45\pms{0.27}$  & $9.14\pms{0.22}$  & $21.77\pms{0.73}$ & $22.88\pms{0.93}$ & $24.27\pms{0.13}$ & $25.16\pms{0.33}$ & $21.14\pms{0.83}$ & $22.15\pms{0.72}$ \\
    $\mathbb{H}$  & $8.59\pms{0.21}$  & $9.36\pms{0.18}$  & $21.95\pms{0.88}$ & $23.21\pms{0.83}$ & $25.11\pms{0.15}$ & $26.00\pms{0.22}$ & $\textbf{26.87}\pms{0.38}$ & $\textbf{27.80}\pms{0.34}$ \\
    $\mathbb{O}$  & $\textbf{8.71}\pms{0.23}$ & $\textbf{9.54}\pms{0.24}$ & $\textbf{22.13}\pms{0.59}$ & $\textbf{23.71}\pms{0.61}$  & $\textbf{25.40}\pms{0.21}$ & $\textbf{26.30}\pms{0.27}$ & $26.53\pms{0.41}$ & $27.62\pms{0.28}$ \\
    $\mathbb{S}$  & $8.14\pms{0.34}$  & $9.35\pms{0.13}$  & $21.23\pms{0.78}$ & $22.47\pms{0.73}$ & $25.21\pms{0.16}$ & $26.25\pms{0.23}$  & $25.92\pms{0.67}$ & $26.97\pms{0.33}$ \\
    \bottomrule
    \end{tabular}%
  \label{tab: hyper_types_map}%
  \end{threeparttable}
\end{table}%

\subsection{Impact of Hypercomplex Algebra Types}
To investigate the impact of different hypercomplex algebras on predictive performance, we evaluate four types of hypercomplex numbers—complex numbers ($\mathbb{C}$), quaternions ($\mathbb{H}$), octonions ($\mathbb{O}$), and sedenions ($\mathbb{S}$)—with results presented in Table \ref{tab: hyper_types_hits} and Table \ref{tab: hyper_types_map}. The results demonstrate that (\textbf{O1}) hypercomplex based methods generally outperform baseline approaches in the real-valued domain, confirming the superior capability of hypercomplex algebras in modeling complex user interaction patterns. (\textbf{O2}) As the dimensionality of hypercomplex algebras increases, model performance may degrade. This could stem from the trade-off between the algebraic dimension ($2^{m+1}$) and the component embedding size ($d' = d / 2^{m+1}$) under a fixed embedding size ($d$), where an overly small $d'$ may compromise the model’s representational capacity. (\textbf{O3}) Given the variability in user interaction patterns across datasets, the optimal hypercomplex configuration should be tailored to specific contexts. Quaternions ($\mathbb{H}$) exhibit stable performance in most scenarios, making them a reliable choice. Consequently, experiments in this study utilize quaternion-based results, which still offer potential for further performance enhancement.

\begin{table}[!htb]\setlength{\tabcolsep}{0.9mm}
  \centering
  \newcommand{\pms}[1]{\pm \text{\scriptsize #1}}
  \begin{threeparttable}
  \caption{Ablation study on Hits@50 and MAP@50 (\%).}
    \begin{tabular}{c|cc|cc|cc|cc}
    \toprule
    \multicolumn{1}{c|}{Dataset} & \multicolumn{2}{c|}{Android} & \multicolumn{2}{c|}{Christianity} & \multicolumn{2}{c|}{Twitter} & \multicolumn{2}{c}{Memetracker} \\
    \midrule  
    @50 & Hits & MAP & Hits & MAP & Hits & MAP & Hits & MAP \\
    \midrule
    w/o ODE & $31.98\pms{0.31}$    & $9.06\pms{0.08}$  & $58.92\pms{0.63}$ & $\textbf{23.25}\pms{0.11}$ & $57.15\pms{0.53}$  & $20.07\pms{0.49}$ & $65.06\pms{0.25}$ & $26.12\pms{0.14}$ \\
    w/o GNN & $25.50\pms{1.52}$  & $7.86\pms{0.94}$  & $52.45\pms{0.46}$ & $19.92\pms{0.86}$ & $49.66\pms{1.23}$ & $17.80\pms{1.01}$ & $58.34\pms{0.28}$ & $22.40\pms{0.21}$ \\
    w/o HT & $26.54\pms{0.53}$ & $7.83\pms{0.39}$  & $53.45\pms{0.68}$ & $20.17\pms{0.41}$  & $50.10\pms{0.56}$  & $19.83\pms{0.61}$ & $63.46\pms{0.53}$ & $22.31\pms{0.26}$ \\
    w/o LM & $32.06\pms{0.55}$ & $9.13\pms{0.12}$  & $52.61\pms{0.73}$ & $21.46\pms{0.45}$ & $50.61\pms{0.87}$ & $18.11\pms{0.83}$ & $61.65\pms{0.69}$ & $22.96\pms{0.35}$ \\
    w/o IM & $23.36\pms{0.73}$ & $8.12\pms{0.58}$  & $51.33\pms{0.42}$ & $22.67\pms{0.23}$ & $54.14\pms{0.91}$ & $24.64\pms{0.43}$ & $65.84\pms{0.33}$ & $26.65\pms{0.26}$ \\
    w/ GAT & $22.26\pms{0.70}$ & $7.31\pms{0.06}$ & $54.02\pms{0.91}$ & $20.92\pms{0.37}$ & $54.86\pms{0.21}$ & $18.83\pms{0.68}$ & $63.37\pms{0.68}$ & $24.25\pms{0.75}$ \\
    w/ Real & $31.81\pms{0.31}$ & $8.83\pms{0.30}$ & $61.09\pms{2.19}$ & $22.42\pms{0.14}$ & $58.40\pms{0.20}$ & $25.15\pms{0.07}$ & $66.19\pms{0.21}$ & $26.19\pms{0.63}$ \\
    CasHyper & $\textbf{33.39}\pms{0.42}$ & $\textbf{9.36}\pms{0.18}$ & $\textbf{64.06}\pms{1.42}$ & $23.21\pms{0.83}$ & $\textbf{59.20}\pms{0.11}$ & $\textbf{26.00}\pms{0.22}$ & $\textbf{67.29}\pms{0.57}$ & $\textbf{27.80}\pms{0.34}$ \\
    \bottomrule
    \end{tabular}%
  \label{tab: ablation}%
  \end{threeparttable}
\end{table}%

\begin{figure}
\centering
\includegraphics[width=0.8\textwidth]{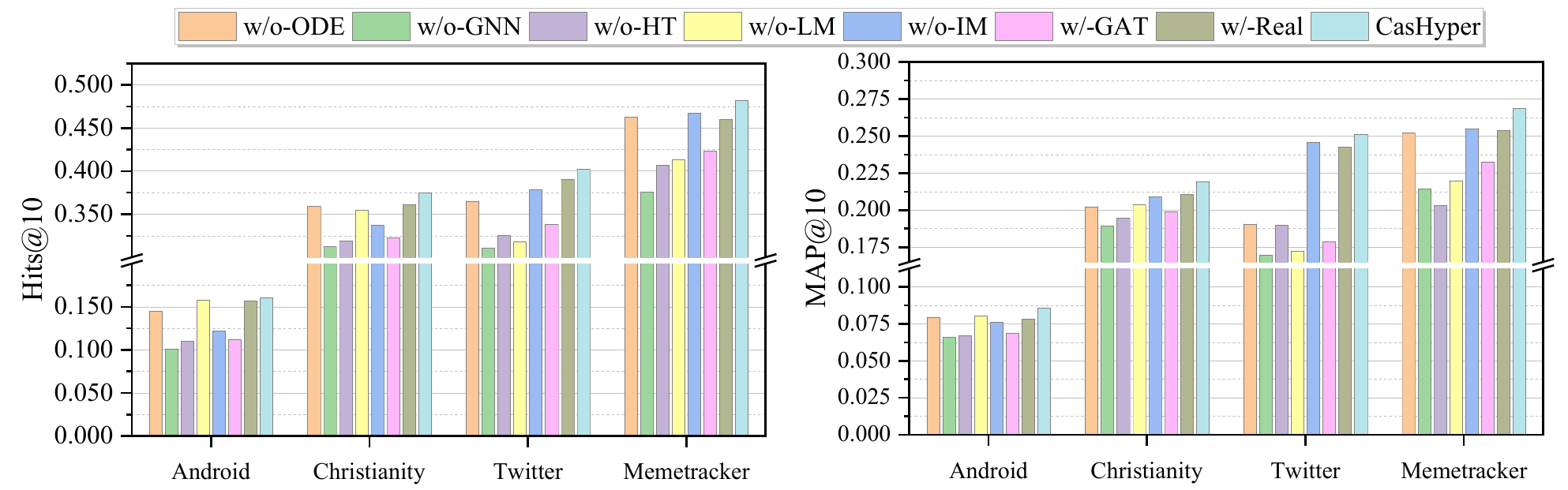}
\caption{The results of ablation study on Hits@10 and MAP@10.}
\label{fig:ablation}
\end{figure}

\subsection{Ablation Study}
To examine the contributions of the major components and design choices in CasHyper, we construct seven variants. (1) \textbf{w/o-ODE} removes the continuous dynamic encoding channel based on Neural ODEs; (2) \textbf{w/o-GNN} removes the static graph encoding channel; (3) \textbf{w/o-HT} replaces the hypercomplex Transformer with a conventional real-valued Transformer equipped with relative positional encoding; (4) \textbf{w/o-LM} removes the linear prediction module and retains only the interaction-aware module; and (5) \textbf{w/o-IM} removes the interaction-aware prediction module and relies on the linear prediction module. In addition, we introduce two comparison variants in response to the architectural design questions. (6) \textbf{w/-Real} replaces the hypercomplex operations in CasHyper with real-valued operations while retaining the overall architecture, embedding dimensionality, and a comparable number of learnable parameters. (7) \textbf{w/-GAT} replaces the GCN neighborhood aggregation with a graph attention mechanism that learns node-specific weights for neighboring nodes, while keeping the remaining modules unchanged.

Table \ref{tab: ablation} and Figure \ref{fig:ablation} report the results on Hits@50 and MAP@50, and Hits@10 and MAP@10, respectively. Overall, CasHyper achieves the strongest or near-strongest performance across the evaluated datasets and metrics. In Table~\ref{tab: ablation}, it obtains the best results on seven of the eight reported metrics. The only exception is MAP@50 on Christianity, where w/o-ODE achieves $23.25\%$, marginally higher than the $23.21\%$ obtained by CasHyper. The following observations can be made:

(\textbf{O1}) Both structural and continuous dynamic encodings contribute to cascade prediction. Removing the static graph encoder leads to clear performance degradation across all datasets, indicating the importance of historical user-user and user-cascade structural information. Removing the Neural ODE channel also generally decreases performance, although its effect varies across datasets and metrics. This suggests that static graph structure provides the primary structural basis, while continuous dynamic encoding supplies complementary information about evolving node and edge states.

(\textbf{O2}) Replacing the hypercomplex Transformer with a conventional real-valued Transformer generally reduces performance, indicating that the proposed sequence encoder contributes to modeling direction-sensitive and cross-component interaction patterns. However, this result should be interpreted as empirical evidence under the current experimental settings rather than as proof that hypercomplex representations universally outperform real-valued sequence models.

(\textbf{O3}) The w/-Real variant provides a controlled comparison between hypercomplex and real-valued implementations. Although w/-Real retains the main architecture and a comparable parameter count, CasHyper performs better on all eight Hits@50 and MAP@50 comparisons. For example, CasHyper improves Hits@50 over w/-Real from $31.81\%$ to $33.39\%$ on Android, from $61.09\%$ to $64.06\%$ on Christianity. These results suggest that the gains cannot be attributed solely to the overall architecture or parameter scale and provide empirical support for the use of structured hypercomplex representations. 

(\textbf{O4}) The w/-GAT variant does not improve upon CasHyper on the evaluated datasets. Although graph attention allows the model to learn different importance weights for neighboring nodes, its performance is consistently lower than that of the aggregation used in CasHyper. One possible explanation is that the learned attention weights may introduce additional estimation difficulty or amplify noisy neighbor relations in the heterogeneous graph. The results indicate that, under the current architecture and datasets, explicitly introducing neighbor-level attention is not necessarily more effective than GCN-based aggregation.

(\textbf{O5}) The two prediction modules exhibit complementary behavior. Removing either the linear module or the interaction-aware module leads to performance degradation on most datasets. The interaction-aware module is particularly important for capturing relations between cascade representations and candidate-user embeddings, while the linear module provides an additional prediction signal based on the encoded historical representation. Their combination generally produces more stable performance across datasets.

\begin{figure}
\centering
\includegraphics[width=\textwidth]{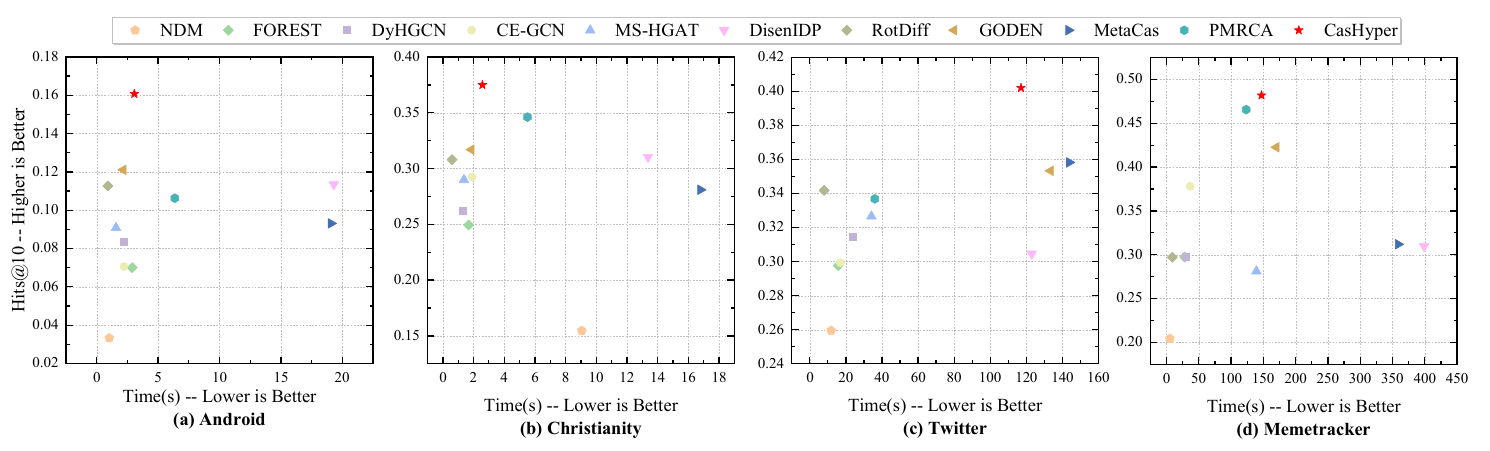}
\caption{Comparison of Hits@10 and runtime on the four datasets.}
\label{fig:run_time}
\end{figure}

\subsection{Efficiency Analysis}
We compare the training efficiency of CasHyper with the baseline methods under the same hardware environment to evaluate its efficiency. All methods are executed on a platform equipped with a 3.9 GHz 20-core Intel processor, 32 GB of system memory, and an NVIDIA GeForce RTX 5070 Ti GPU with 16 GB of video memory. The reported runtime is the average training time per epoch across multiple independent runs.

Figure~\ref{fig:run_time} jointly presents Hits@10 and the average training time per epoch, where higher predictive performance and lower training time are preferred. CasHyper achieves the highest Hits@10 on all four datasets, but its computational efficiency varies with dataset scale.
On Android and Christianity, CasHyper requires only a few seconds per epoch and remains competitive with, or faster than, several baseline methods. This suggests that the additional hypercomplex computations are manageable on relatively small datasets.
Twitter contains the largest number of social links, while Memetracker contains the largest number of cascades. On these two datasets, CasHyper requires noticeably more training time than several baselines, indicating that the current framework has limited scalability when the graph structure or cascade collection becomes large. The additional cost mainly arises from the hypercomplex operations and the rotation process.

These findings highlight an important limitation of the current framework. Future work may improve scalability through more efficient hypercomplex operations, sparse neighborhood sampling, simplified ODE solvers, mini-batch graph training, or lightweight sequence encoders.

\begin{figure*}
\centering
\includegraphics[width=0.98\textwidth]{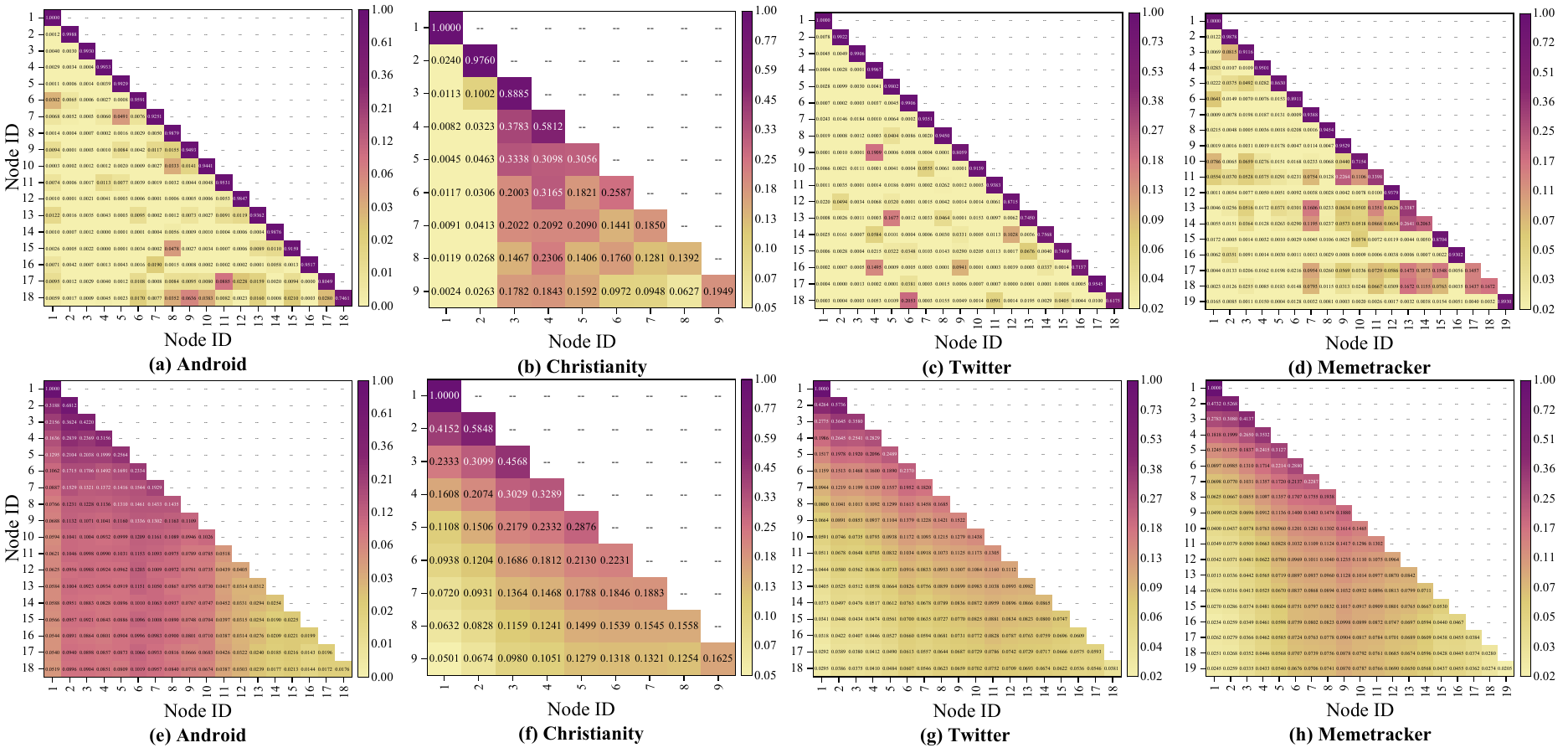}
\caption{Attention coefficient matrices computed by the standard self-attention mechanism (top) and the hypercomplex rotation-based attention mechanism (bottom) on randomly selected cascades.}
\label{fig:attention}
\end{figure*}

\subsection{Parameter Analysis}
\subsubsection{Visualization and Quantitative Analysis for Relative Positional Modeling}
To verify the effectiveness of hypercomplex rotational transformations in capturing relative positional relationships among nodes within a cascade, we visualize the attention coefficient matrices computed by the standard self-attention mechanism, as defined in Eq. (\ref{eq:self_attention}), and the hypercomplex rotation-based attention mechanism, as defined in Eq. (\ref{eq:rotation_att_score}), respectively. Figure \ref{fig:attention} presents the pairwise attention distributions of nodes within a randomly selected cascade from different datasets.

As shown in the figure, the standard self-attention mechanism primarily concentrates attention weights along the diagonal, indicating a strong bias toward assigning higher importance to each node itself, while exhibiting limited capability in modeling sequential order and relative positional relationships among nodes. In contrast, the attention mechanism equipped with hypercomplex rotational transformations explicitly encodes relative positional information, thereby assigning higher attention weights to nodes that appear earlier in the cascade and exhibit stronger interactions between nodes. These observations illustrate that hypercomplex rotations encourage the attention mechanism to incorporate more historical and relative positional information within cascade sequences.

We further conduct a quantitative analysis on the test set. Specifically, we introduce two attention-distribution metrics, namely historical attention ratio (HAR) and self-diagonal ratio (SDR), to characterize how attention weights are distributed over historical positions and self-positions. For an attention matrix $\mathbf{A} \in \mathbb{R}^{L_c \times L_c}$ of a cascade $C_c$, where $A_{ij}$ denotes the attention weight from the $i$-th position to the $j$-th position, 
\begin{equation}
\nonumber
    \mathrm{HAR}=\frac{1}{L_c-1}
    \sum_{i=2}^{L_c}\frac{\sum_{j<i} A_{ij}}{ \sum_{j\leq i} A_{ij}}, \quad
    \mathrm{SDR}=\frac{1}{L_c-1}
    \sum_{i=2}^{L_c}\frac{A_{ii}}{\sum_{j\leq i} A_{ij}}.
\end{equation}
HAR measures the proportion of attention assigned to previously activated nodes. A higher HAR value indicates that the model attends more to historical cascade contexts. In contrast, SDR measures the concentration of attention weights on the diagonal self-position. A higher SDR value indicates stronger self-position dominance, suggesting that the attention mechanism tends to focus on each node itself rather than on other historical nodes. Since the first position has no historical context, both HAR and SDR are averaged from the second position.

As shown in Table~\ref{tab:har_sdr_analysis}, the standard self-attention mechanism exhibits high SDR values and relatively low HAR values across the four datasets, indicating that its attention distribution is largely dominated by self-position attention. This finding is consistent with the diagonal-dominant patterns observed in Figure~\ref{fig:attention}.
In contrast, the proposed hypercomplex rotation-based attention mechanism consistently achieves higher HAR values and lower SDR values. This indicates that the proposed mechanism alleviates excessive self-position concentration and assigns more attention to previously activated nodes in the cascade. These results provide quantitative evidence that hypercomplex rotational transformations help the attention mechanism better capture historical dependency patterns and relative positional structures in cascade sequences.
Together, the case visualization and the test-set-level quantitative results provide evidence that the proposed mechanism reduces excessive self-position concentration and strengthens the use of historical cascade contexts.

\begin{table*}[!htbp]
\centering
\newcommand{\pms}[1]{\pm \text{\scriptsize #1}}
\caption{Quantitative comparison of attention distributions on the test set (\%). SA denotes the standard self-attention mechanism, and HA denotes the hypercomplex rotation-based attention mechanism.}
\label{tab:har_sdr_analysis}
\renewcommand{\arraystretch}{1.15}
\begin{tabular}{lcccccccc}
\toprule
\multirow{2}{*}{Type}
& \multicolumn{2}{c}{Android}
& \multicolumn{2}{c}{Christianity}
& \multicolumn{2}{c}{Twitter}
& \multicolumn{2}{c}{Memetracker} \\
\cmidrule(lr){2-3}
\cmidrule(lr){4-5}
\cmidrule(lr){6-7}
\cmidrule(lr){8-9}
& HAR & SDR & HAR & SDR & HAR & SDR & HAR & SDR \\
\midrule
SA
& $4.88\pms{0.23}$ & $95.52\pms{0.22}$
& $17.03\pms{3.22}$ & $85.29\pms{2.78}$
& $15.85\pms{2.51}$ & $86.24\pms{2.05}$
& $2.75\pms{0.61}$ & $97.41\pms{0.56}$ \\
HA
& $\mathbf{81.48}\pms{1.85}$ & $\mathbf{25.32}\pms{1.69}$
& $\mathbf{66.64}\pms{2.38}$ & $\mathbf{43.14}\pms{2.04}$
& $\mathbf{79.20}\pms{3.54}$ & $\mathbf{33.24}\pms{2.91}$
& $\mathbf{73.57}\pms{0.96}$ & $\mathbf{32.25}\pms{0.78}$ \\
\bottomrule
\end{tabular}
\end{table*}

\begin{figure}
\centering
\includegraphics[width=0.8\textwidth]{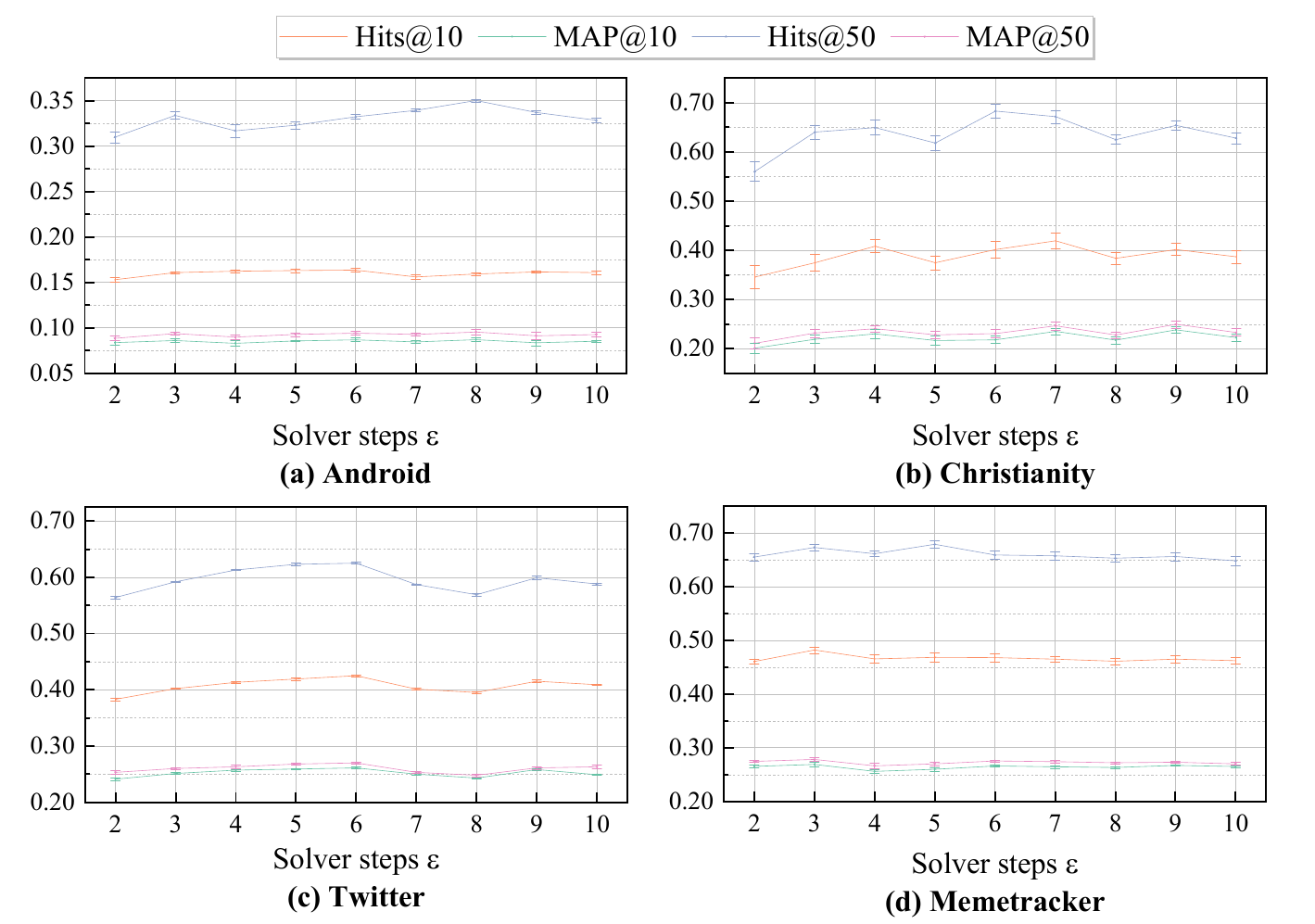}
\caption{ Impact of solver steps $\epsilon$ on model performance.}
\label{fig:varying_t}
\end{figure}

\subsubsection{Impact of Solver Steps}
In Neural ODEs, the evolution of model states is realized through numerical solvers, where the number of solver steps $\epsilon$ controls the granularity of the numerical simulation. To investigate the impact of solver-step selection on model performance, we vary $\epsilon$ from 2 to 10 on four datasets, with the results summarized in Figure \ref{fig:varying_t}. The model exhibits remarkable robustness. Across all evaluated scenarios, the model sustains competitive predictive performance even with a minimal step count of $\epsilon=2$. As $\epsilon$ increases, the performance may improve slightly on some datasets or metrics, indicating that a finer numerical simulation can sometimes better capture continuous temporal dynamics. However, further increasing $\epsilon$ also introduces additional computational cost and may lead to slight fluctuations, possibly because overly fine-grained temporal discretization can amplify noise or inherent uncertainties in the data. Therefore, we set $\epsilon=3$ as the default solver-step configuration as a practical trade-off between predictive performance, robustness, and computational efficiency.

\begin{figure}
\centering
\includegraphics[width=\textwidth]{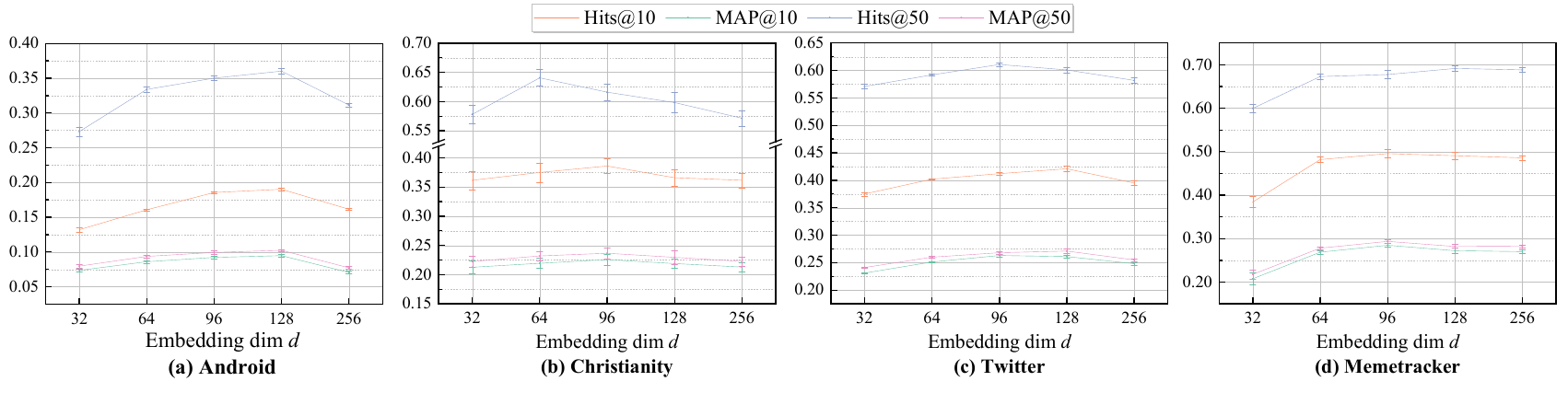}
\caption{ Impact of embedding dimension $d$ on model performance.}
\label{fig:varying_dim}
\end{figure}

\subsubsection{Impact of Embedding Dimension}
In our framework, increasing $d$ scales the dimension of hypercomplex sub-units, potentially enhancing the model's ability to capture complex inter-component correlations. We analyze the impact of $d \in \{32, 64, 96, 128, 256\}$ on model performance. As illustrated in Figure \ref{fig:varying_dim}, the performance across all metrics initially exhibits a significant upward trend as $d$ increases. This suggests that a larger embedding dimension in hypercomplex space provides a greater representational degree of freedom, enabling the model to capture latent propagation patterns more effectively. However, once the performance reaches its peak, further increasing $d$ results in a noticeable decline. This phenomenon indicates that while expanding  the dimension enhances model capacity, an excessively high $d$ may trigger overfitting, especially given the structured interaction constraints inherent in hypercomplex algebra. Larger embedding dimensions also introduce higher computational and memory costs. Considering the trade-off between performance gain and computational efficiency, we select $d=64$ as the default total embedding dimension.

\begin{table}[!htb]\setlength{\tabcolsep}{0.8mm} 
  \centering
  \newcommand{\pms}[1]{\pm \text{\scriptsize #1}}
  \begin{threeparttable}
  \caption{Results of different $\lambda_{\text{cons}}$ on Hits Metric (\%).}
    \begin{tabular}{l|cc|cc|cc|cc}
    \toprule
    \multirow{2}[5]{*}{$\lambda_{\text{cons}}$} & \multicolumn{2}{c|}{Android}  & \multicolumn{2}{c|}{Christianity} & \multicolumn{2}{c|}{Twitter}   & \multicolumn{2}{c}{Memetracker} \\
\cmidrule{2-9}          & @10    & \multicolumn{1}{c|}{@50}    & @10    & \multicolumn{1}{c|}{@50}    & @10    & \multicolumn{1}{c|}{@50}    & @10    & \multicolumn{1}{c}{@50} \\
    \midrule
    $0.6$ & $15.89\pms{0.03}$ & $31.87\pms{0.35}$  & $36.83\pms{1.56}$ & $62.49\pms{1.57}$  & $39.59\pms{0.61}$  & $58.80\pms{0.38}$ & $44.28\pms{0.48}$ & $65.37\pms{0.52}$ \\
    $0.8$ & $15.54\pms{0.25}$ & $32.34\pms{0.61}$  & $37.35\pms{1.90}$  & $62.72\pms{1.45}$ & $39.73\pms{0.25}$  & $\mathbf{59.34}\pms{0.35}$  & $44.16\pms{0.30}$ & $65.53\pms{0.24}$ \\
    $1.0$ & $\mathbf{16.07}\pms{0.13}$ & $\mathbf{33.39}\pms{0.42}$  & $37.50\pms{1.65}$  & $64.06\pms{1.42}$ & $\mathbf{40.20}\pms{0.12}$  & $59.20\pms{0.11}$  & $\textbf{48.19}\pms{0.61}$ & $\textbf{67.29}\pms{0.57}$ \\
    $1.2$ & $15.21\pms{0.54}$ & $32.26\pms{0.12}$ & $\mathbf{37.95}\pms{2.46}$ & $\mathbf{64.56}\pms{1.90}$ & $39.89\pms{0.72}$ & $58.59\pms{0.65}$ & $46.56\pms{0.69}$ & $66.13\pms{0.79}$ \\
    $1.4$ & $15.62\pms{0.12}$ & $32.07\pms{0.62}$  & $36.05\pms{1.00}$  & $61.38\pms{2.68}$ & $39.60\pms{0.32}$ & $58.71\pms{0.07}$ & $45.43\pms{2.58}$ & $65.05\pms{1.98}$ \\
    \bottomrule
    \end{tabular}%
  \label{tab: lambda_hits}%
  \end{threeparttable}
\end{table}%

\begin{table}[!htb]\setlength{\tabcolsep}{0.8mm} 
  \centering
  \newcommand{\pms}[1]{\pm \text{\scriptsize #1}}
  \begin{threeparttable}
  \caption{Results of different $\lambda_{\text{cons}}$ on MAP Metric (\%).}
    \begin{tabular}{l|cc|cc|cc|cc}
    \toprule
    \multirow{2}[5]{*}{$\lambda_{\text{cons}}$} & \multicolumn{2}{c|}{Android}  & \multicolumn{2}{c|}{Christianity} & \multicolumn{2}{c|}{Twitter}   & \multicolumn{2}{c}{Memetracker} \\
\cmidrule{2-9}          & @10    & \multicolumn{1}{c|}{@50}    & @10    & \multicolumn{1}{c|}{@50}    & @10    & \multicolumn{1}{c|}{@50}   & @10    & \multicolumn{1}{c}{@50} \\
    \midrule
    $0.6$  & $8.37\pms{0.18}$  & $9.23\pms{0.07}$  & $21.17\pms{0.57}$ & $22.19\pms{0.59}$ & $24.91\pms{0.14}$ & $25.78\pms{0.16}$ & $23.99\pms{0.75}$ & $25.01\pms{0.74}$ \\
    $0.8$  & $8.42\pms{0.23}$  & $9.23\pms{0.19}$  & $21.57\pms{0.50}$ & $22.63\pms{0.96}$ & $\mathbf{25.25}\pms{0.05}$ & $\mathbf{26.15}\pms{0.05}$ & $23.76\pms{0.61}$ & $24.80\pms{0.61}$ \\
    $1.0$  & $\mathbf{8.59}\pms{0.21}$  & $\mathbf{9.36}\pms{0.18}$  & $\mathbf{21.95}\pms{0.88}$ & $\mathbf{23.21}\pms{0.83}$ & $25.11\pms{0.15}$ & $26.00\pms{0.22}$ & $\textbf{26.87}\pms{0.38}$ & $\textbf{27.80}\pms{0.34}$ \\
    $1.2$  & $8.53\pms{0.17}$ & $9.29\pms{0.13}$ & $21.72\pms{1.23}$ & $22.72\pms{1.20}$  & $25.05\pms{0.12}$ & $25.93\pms{0.32}$ & $25.48\pms{0.44}$ & $26.52\pms{0.45}$ \\
    $1.4$  & $8.49\pms{0.21}$  & $9.26\pms{0.21}$  & $21.02\pms{0.34}$ & $22.17\pms{0.42}$ & $25.01\pms{0.21}$ & $25.51\pms{0.18}$  & $23.75\pms{2.58}$ & $24.80\pms{2.01}$ \\
    \bottomrule
    \end{tabular}%
  \label{tab: lambda_map}%
  \end{threeparttable}
\end{table}%

\subsubsection{Impact of Weighting Coefficient $\lambda_{\text{cons}}$}
The weighting coefficient $\lambda_{\mathrm{cons}}$ controls the relative contribution of the consistency regularization term to the overall training objective. To investigate its influence, we vary $\lambda_{\mathrm{cons}}$ within $\{0.6, 0.8, 1.0, 1.2, 1.4\}$ and report the Hits and MAP results in Tables~\ref{tab: lambda_hits} and~\ref{tab: lambda_map}, respectively. Overall, CasHyper exhibits relatively stable performance across the tested values, particularly when $\lambda_{\mathrm{cons}}$ lies between $0.8$ and $1.2$. Among these settings, $\lambda_{\mathrm{cons}}=1.0$ achieves the best results on all Android and Memetracker metrics, as well as on most MAP metrics and several Hits metrics on the other datasets. Although $\lambda_{\mathrm{cons}}=0.8$ or $1.2$ yields slightly better performance on a few individual metrics, the differences are generally modest, and $\lambda_{\mathrm{cons}}=1.0$ provides the most consistently competitive performance across datasets and evaluation metrics. When $\lambda_{\mathrm{cons}}$ is relatively small, the regularization signal may be insufficient to effectively encourage the alignment of the query and key rotary parameters. In contrast, an excessively large value may place too much emphasis on parameter consistency and interfere with the optimization of the primary prediction objective. Therefore, we set $\lambda_{\mathrm{cons}}=1.0$ as the default configuration, as it provides a practical balance between predictive performance and the consistency constraint across the evaluated datasets.

\begin{figure}
\centering
\includegraphics[width=\textwidth]{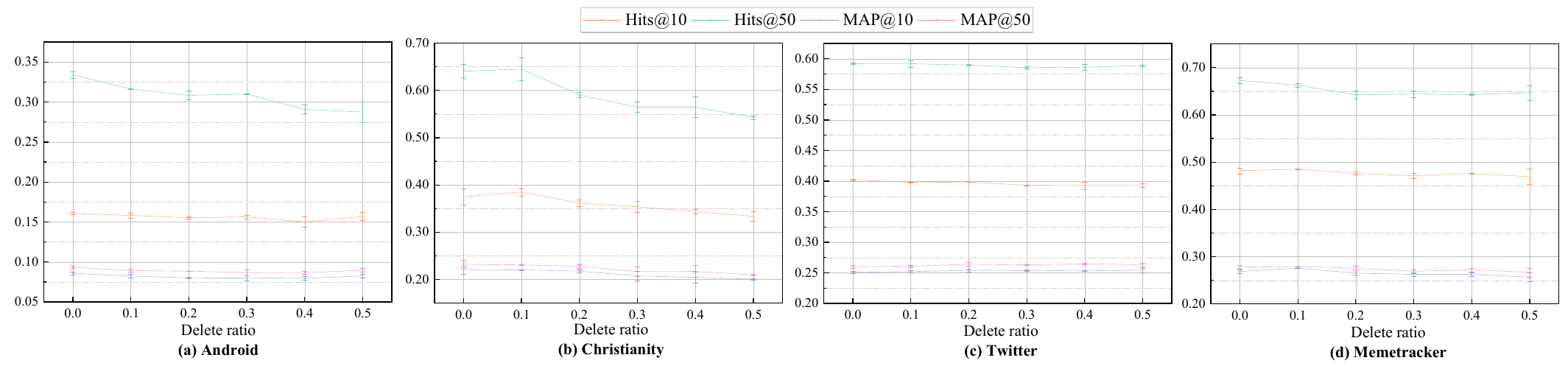}
\caption{Robustness of CasHyper under different random edge deletion ratios. Error bars indicate the standard deviations.}
\label{fig:del_ratio}
\end{figure}

\subsubsection{Robustness to Incomplete Graph Structures}
To examine the robustness of CasHyper under structural incompleteness, we conduct an additional perturbation experiment by randomly removing a proportion of edges from the original input graph. Specifically, the edge deletion ratio is varied from $0$ to $0.5$ with an interval of $0.1$. For each deletion ratio, edges are sampled uniformly at random and removed before model training. The remaining model configuration and training protocol are kept unchanged.

Figure~\ref{fig:del_ratio} presents the results across the four datasets. Overall, CasHyper maintains relatively stable performance under moderate levels of edge removal, although the degree of sensitivity differs across datasets.
On Android, the performance decreases gradually as more edges are removed. The decline is more evident for Hits@50, whereas Hits@10 and the MAP metrics remain comparatively stable. This indicates that structural incompleteness has a greater effect on retrieving a broader set of candidate users than on identifying the most likely next participants.
Christianity is more sensitive to missing graph structure. Its performance remains relatively stable under a small deletion ratio but declines more clearly when the deletion ratio exceeds $0.2$. For example, Hits@50 decreases from approximately $0.64$ to $0.54$ when half of the edges are removed. This suggests that the prediction of Christianity depends more strongly on the availability of graph connectivity.
By contrast, Twitter exhibits the strongest robustness. Even at a deletion ratio of $0.5$, all four metrics fluctuate only slightly, with Hits@50 remaining close to $0.59$. Memetracker also shows relatively stable behavior under moderate edge deletion, although the performance variation and standard deviation become somewhat larger at the highest deletion ratio.

These results suggest that CasHyper does not rely exclusively on complete graph connectivity. One possible reason is that its prediction is based on multiple complementary information sources. The static neighborhood aggregation captures the observed structural relations, whereas the continuous asymmetric encoding and the hypercomplex-based interaction model evolve interactions and historical activation patterns. Consequently, when part of the graph structure is missing, the remaining channels can partially compensate for the lost neighborhood information.
Nevertheless, the results also show that robustness is dataset-dependent, and performance generally decreases as more edges are removed. Therefore, this experiment should be interpreted as evidence of resilience to randomly missing structural links rather than a guarantee of robustness to arbitrary graph noise. More challenging settings, such as spurious-edge injection, non-random missingness, attribute corruption, and adversarial graph perturbations, remain directions for future investigation.

\section{Conclusion and Discussion}
\label{section:conclusion}
In this paper, we revisit the challenge of information cascade prediction from a novel algebraic perspective, addressing the inherent limitations of conventional real-valued Euclidean embeddings in capturing multi-factor interactions and asymmetric social influences. To this end, we introduce CasHyper, a principled framework that harmonizes hypercomplex algebra with Graph Neural Networks (GNNs) and Transformers. In this model, node representations are organized into hypercomplex components, static neighborhood information is aggregated over the heterogeneous graph, continuous dynamic edge evolution is modeled through a Neural ODE-based channel, and cascade sequences are further encoded using a hypercomplex Transformer with rotary positional transformations. The experimental results across four real-world datasets show that CasHyper achieves competitive or better performance than the compared baselines.   In addition, the analysis of various Cayley-Dickson algebras suggests that hypercomplex structures, such as quaternions and octonions, may exhibit distinct empirical behaviors across datasets. These results suggest that hypercomplex representations can be a promising modeling choice for cascade prediction, especially when user interactions involve multiple latent factors and direction-sensitive dependencies.

From a methodological perspective, CasHyper provides a structured way to organize node embeddings into multiple algebraic components and to use hypercomplex products for constructing order-dependent interaction representations. The non-commutativity of higher-dimensional hypercomplex multiplication enables the model to distinguish the representation of an ordered pair $(u_i,u_j)$ from that of $(u_j,u_i)$ prior to the neural mapping, thereby enabling it to learn asymmetric influence patterns from data. The hypercomplex Transformer further incorporates rotary positional transformations into the cascade sequence encoder. Since the rotation operation preserves the norms of the transformed vectors, positional information can be introduced while maintaining the magnitudes of the representations. 

From a practical perspective, CasHyper provides a structured approach for modeling information cascade prediction on social platforms by organizing node representations into multiple hypercomplex components. Under the evaluated benchmark settings, the model shows competitive predictive performance. These results suggest that hypercomplex representations may be useful for capturing complex interaction patterns in cascade prediction tasks.
The practical relevance of this work should be understood within the scope of predictive modeling. More accurate prediction of the next participant in an information cascade can potentially serve as a supporting component for downstream applications such as diffusion analysis, early warning systems, and platform-level decision support. Applying the proposed model to these scenarios would require additional task-specific evaluation protocols and deployment-oriented analyses. Therefore, we view CasHyper as a prediction-oriented modeling framework that may support such applications in future work, rather than as a complete solution for these broader practical scenarios.

A limitation of the current framework lies in its sensitivity to data sparsity in the early stages of a cascade. Since CasHyper relies on capturing multi-factor interactions, its performance may be constrained when the initial diffusion observations are extremely limited. Future research will explore the integration of self-supervised learning or contrastive learning objectives to enhance the robustness of hypercomplex embeddings under low-resource conditions. Furthermore, investigating the interpretability of the imaginary basis components could provide deeper insights.

\appendix
\section{Discussion of Asymmetry-aware Representation}
\label{appendix:asymmetric influence}
According to the model definition, the influence from node $u_i$ to node $u_j$ and that from node $u_j$ to node $u_i$ are computed as,
\begin{equation} \nonumber 
\frac{d w_{e_{ij}}^{(t)}}{dt} = f(\mathbf{X}_{u_i}^{(t)} \otimes_{m+1} \mathbf{X}_{u_j}^{(t)}), \quad \frac{d w_{e_{ji}}^{(t)}}{dt} = f(\mathbf{X}_{u_j}^{(t)} \otimes_{m+1} \mathbf{X}_{u_i}^{(t)}). 
\end{equation} 
A fundamental property of the Cayley-Dickson construction is that, for $m \geq 1$, the resulting multiplication operator $\otimes_{m+1}$ is generally non-commutative. Therefore, for two node states $\mathbf{X}_{u_i}^{(t)}$ and $\mathbf{X}_{u_j}^{(t)}$, their multiplicative interaction is order-dependent in general case, 
\begin{equation} \nonumber
\mathbf{X}_{u_i}^{(t)} \otimes_{m+1} \mathbf{X}_{u_j}^{(t)} \neq \mathbf{X}_{u_j}^{(t)} \otimes_{m+1} \mathbf{X}_{u_i}^{(t)}.
\end{equation}
This order dependence provides different directional interaction representations before they are passed to the neural mapping $f$. We note that non-commutativity alone does not strictly guarantee different scalar outputs after the mapping $f$, since the final outputs also depend on the learned parameters and the specific input states. Nevertheless, by preserving the order information in the algebraic interaction term, the proposed encoding provides a structural inductive bias that enables the model to learn asymmetric influence patterns when the data support them. Thus, the proposed continuous asymmetric encoding does not impose asymmetric outputs in all cases, but equips the model with the representational capacity to distinguish the influence from $u_i$ to $u_j$ and that from $u_j$ to $u_i$.

\section{Explanation of Property \ref{property:norm preservation}}
\label{appendix:norm preservation}
\begin{definition} [\textnormal{Hypercomplex Rotational Transformation}]
Here, $\mathbf{v}$ consists of $2^{m+1}$ real-valued component vectors, and $\mathbf{v}_k$ denotes the coefficient vector corresponding to the $k$-th hypercomplex basis component. For an arbitrary input $\mathbf{v} \in \mathbb{R}^{2^{m+1} \times d'}$, the rotational transformation $\mathcal{R}$ on its $k$-th hypercomplex component $\mathbf{v}_k \in \mathbb{R}^{d'}$ at an absolute position $pos$ is defined as,
\begin{equation}
\tilde{\mathbf{v}}_k =
\mathcal{R}(\mathbf{v}_k) = \mathbf{v}_k \cdot e_k^{\mathbf{i}(2\pi \bm{\omega}_k
\text{pos}\  +\  \bm{\theta}_k)},
\nonumber
\end{equation}
where $\mathbf{i}$ is a unit hypercomplex imaginary number satisfying $\mathbf{i}^2 = -1$, and $\bm{\omega}_k$ and $\bm{\theta}_k$ are learnable frequency and phase parameters dynamically generated from the input vector.
\label{definition:Hypercomplex Rotational}
\end{definition}

According to Definition \ref{definition:Hypercomplex Rotational}, the transformed vector component can be written as 
$\tilde{\mathbf{v}}_k = \mathbf{v}_k \cdot e^{\mathbf{i}\phi_k}$, where 
$\phi_k = 2\pi \bm{\omega}_k \text{pos} + \bm{\theta}_k$. 
The factor $e^{\mathbf{i}\phi_k}$ is a unit-modulus rotational operator. 
By Euler's formula, we have $e^{\mathbf{i}\phi_k}=\cos(\phi_k)+\mathbf{i}\sin(\phi_k)$.
Thus, the transformed component can be expanded as $\tilde{\mathbf{v}}_k = \mathbf{v}_k\cos(\phi_k) + \mathbf{i}\mathbf{v}_k\sin(\phi_k)$.
Here, $\mathbf{i}$ denotes the imaginary basis direction. 
When represented in the corresponding two real-valued subspace, this transformation maps each entry of $\mathbf{v}_k$ to a rotated pair $\left(\mathbf{v}_k\cos(\phi_k), \mathbf{v}_k\sin(\phi_k) \right)$. Hence, the squared L2-norm of the transformed representation is,
\begin{equation}
\nonumber
\|\tilde{\mathbf{v}}_k\|_2^2 = \|\mathbf{v}_k\cos(\phi_k)\|_2^2 + \|\mathbf{v}_k\sin(\phi_k)\|_2^2.
\end{equation}
This yields,
\begin{equation}
\nonumber
\|\tilde{\mathbf{v}}_k\|_2^2 = \cos^2(\phi_k)\|\mathbf{v}_k\|_2^2 + \sin^2(\phi_k)\|\mathbf{v}_k\|_2^2
= \|\mathbf{v}_k\|_2^2.
\end{equation}
Since the L2-norm is non-negative, taking the square root on both sides gives
$\|\tilde{\mathbf{v}}_k\|_2 = \|\mathbf{v}_k\|_2$. We finally obtain
$\|\mathcal{R}(\mathbf{v}_k)\|_2 = \|\mathbf{v}_k\|_2$.
This confirms the norm-preservation property of the rotational transformation.

\section{Explanation of Property~\ref{property:relative position dependence}}
\label{appendix:relative position dependence}
Let $\mathbf{Q}_{j,k},\mathbf{K}_{l,k}\in\mathbb{R}^{d'}$ denote the $k$-th real-valued component vectors of the query and key representations at positions $j$ and $l$, respectively. We define the phase terms as,
\[
\phi^Q_{j,k}=2\pi\omega_k^Q j+\theta_k^Q,\qquad
\phi^K_{l,k}=2\pi\omega_k^K l+\theta_k^K.
\]
After the rotational transformation, the corresponding complex-valued component vectors are given by,
\[
\tilde{\mathbf{Q}}_{j,k}=\mathbf{Q}_{j,k}e^{\mathbf{i}\phi^Q_{j,k}},\qquad
\tilde{\mathbf{K}}_{l,k}=\mathbf{K}_{l,k}e^{\mathbf{i}\phi^K_{l,k}}.
\]
Since $\mathbf{Q}_{j,k}$ and $\mathbf{K}_{l,k}$ are real-valued vectors, the real-valued similarity between the rotated components can be computed as the real part of their Hermitian inner product,
\[
\left\langle \tilde{\mathbf{Q}}_{j,k},\tilde{\mathbf{K}}_{l,k}\right\rangle
=
\mathrm{Re}\left(
\tilde{\mathbf{Q}}_{j,k}^{\top}
\overline{\tilde{\mathbf{K}}_{l,k}}
\right).
\]
Substituting the transformed vectors, we obtain,
\begin{equation}
\begin{aligned}
\left\langle \tilde{\mathbf{Q}}_{j,k},\tilde{\mathbf{K}}_{l,k}\right\rangle
&=
\mathrm{Re}\left(
\left(\mathbf{Q}_{j,k}e^{\mathbf{i}\phi^Q_{j,k}}\right)^{\top}
\overline{
\left(\mathbf{K}_{l,k}e^{\mathbf{i}\phi^K_{l,k}}\right)}
\right) \\
&=\mathrm{Re}\left(
\left(\mathbf{Q}_{j,k}^{\top}\mathbf{K}_{l,k}\right)
e^{\mathbf{i}(\phi^Q_{j,k}-\phi^K_{l,k})}\right) \\
&=\left(\mathbf{Q}_{j,k}^{\top}\mathbf{K}_{l,k}\right)
\cos\left(\phi^Q_{j,k}-\phi^K_{l,k}\right) \\
&=\left(\mathbf{Q}_{j,k}^{\top}\mathbf{K}_{l,k}\right)
\cos\left(2\pi\omega_k^Qj - 2\pi\omega_k^Kl + \theta_k^Q - \theta_k^K \right).
\label{equ:attention score}
\end{aligned}
\end{equation}

\paragraph{Regularization condition.} The above expression becomes explicitly dependent on the relative position $(j-l)$ when the query and key rotary parameters are shared or approximately aligned. In our model, this condition is encouraged by the consistency regularization term $\mathcal{L}_{\mathrm{cons}}$, which penalizes the discrepancy between $(\bm{\omega}^Q,\bm{\theta}^Q)$ and $(\bm{\omega}^K,\bm{\theta}^K)$. This regularization does not strictly prove parameter equality, but provides an optimization bias toward $\bm{\omega}_k^Q \approx \bm{\omega}_k^K$ and $\bm{\theta}_k^Q \approx \bm{\theta}_k^K$. Under this regularization condition, we can write $\bm{\omega}_k^Q \approx \bm{\omega}_k^K = \bm{\omega}_k$ and $\bm{\theta}_k^Q \approx \bm{\theta}_k^K = \bm{\theta}_k$. Then the exponential term in Eq.~\eqref{equ:attention score} is approximately reduced to,
\begin{equation}
\cos \left[ 2\pi(\bm{\omega}_k j-\bm{\omega}_k l) + (\bm{\theta}_k-\bm{\theta}_k) \right] \approx \cos[2\pi\bm{\omega}_k(j-l)]. 
\end{equation}
 Accordingly, the real part of the Hermitian inner product can be approximated as, 
\begin{equation} 
\langle \tilde{\mathbf{Q}}_{j,k}, \tilde{\mathbf{K}}_{l,k} \rangle \approx \operatorname{Re} \left( (\mathbf{Q}_{j,k}^{\top}\mathbf{K}_{l,k}) \cdot e^{\mathbf{i}\cdot 2\pi\bm{\omega}_k(j-l)} \right).
\end{equation}
 This expression shows that, under the regularization condition, the attention score depends on both the semantic interaction $(\mathbf{Q}_{j,k}^{\top}\mathbf{K}_{l,k})$ and the relative distance $(j-l)$. Therefore, the proposed rotary mechanism provides an inductive bias for capturing relative positional information, rather than a strict guarantee that the learned query and key rotary parameters are identical.

\printcredits

% \bibliographystyle{cas-model2-names}
\bibliographystyle{model5-names}

\bibliography{cas-refs}


\end{document}